\documentclass{article}
\usepackage{hyperref}
\usepackage{soul}
\usepackage{graphicx}
\usepackage{placeins}
\title{A 25 Year Retrospective on D-Lib Magazine}

\author{
Michael L. Nelson \\
Old Dominion University \\
Norfolk, VA USA\\
\texttt{mln@cs.odu.edu}\\
\and
Herbert Van de Sompel\\
Data Archiving and Networked Services\\
The Hague, Netherlands\\
\texttt{herbert.van.de.sompel@dans.knaw.nl}\\
}

\begin{document}
\maketitle

\begin{abstract}
\sloppy
In July, 1995 the first issue of D-Lib Magazine was published as an
on-line, HTML-only, open access magazine, serving as the focal point
for the then emerging digital library research community.  In 2017
it ceased publication, in part due to the maturity of the community
it served as well as the increasing availability of and competition
from eprints, institutional repositories, conferences, social media,
and online journals -- the very ecosystem that D-Lib Magazine nurtured
and enabled.  As long-time members of the digital library community
and authors with the most contributions to D-Lib Magazine, we reflect
on the history of the digital library community and D-Lib Magazine,
taking its very first issue as guidance. It contained three articles,
which described: the Dublin Core Metadata Element Set, a project status
report from the NSF/DARPA/NASA-funded Digital Library Initiative (DLI),
and a summary of the Kahn-Wilensky Framework (KWF) which gave us, among
other things, Digital Object Identifiers (DOIs).  These technologies,
as well as many more described in D-Lib Magazine through its 22 years,
have had a profound and continuing impact on the digital library and
general web communities.  

\fussy

\end{abstract}

\section{Introduction}
\label{sec:introduction}

In July, 1995, the Corporation for National Research Initiatives (CNRI)
published the first issue of D-Lib Magazine (\url{www.dlib.org}).
D-Lib Magazine was the most visible and impactful component of the D-Lib
Forum\footnote{\url{http://www.dlib.org/forum/note.html}} administered by
CNRI and funded by the Defense Advanced Research Projects Agency (DARPA).
In July, 2017, D-Lib Magazine published its 265th and final issue,
bringing to a close a successful 22 year run that saw it evolve into an
entity around which the entire digital library (DL) community coalesced.
D-Lib Magazine was itself an innovation: it was published in HTML only and
thereby encouraged exploration in scholarly publishing with hypertext and
hypermedia, it was open access with no article processing charge so it
reached a broad community, its ``magazine'' focus and initially monthly
publication schedule facilitated community building in a pre-blog and
pre-social media world, and it found the elusive middle ground between
researchers and practitioners. \\

During its 22 year run, D-Lib Magazine offered several opportunities
for self-reflection for both the magazine and the community
at large.  In 2000, Bill Arms surveyed the first five years
\cite{arms:2000:editorial}. In 2005, a ten year anniversary special
issue\footnote{\url{https://doi.org/10.1045/july2005-contents}} was
published with contributions from many of the central figures of D-Lib
Magazine and the DL community at large \cite{kahn:2005:editorial,wilson:2005:editorial}.
The 20 year anniversary had a more muted tone, with only the issue's editorial
marking the event \cite{lannom:2015:editorial}; perhaps because editor Larry
Lannom knew the time for the final editorial was not far off \cite{lannom:2017:editorial}.
So we take this, the 25 year anniversary of the first issue,
to reflect on the impact of D-Lib Magazine, both for the information that
it conveyed as well as a proof-of-concept for many DL and web concepts
and technologies that we enjoy today.  We provide this retrospective as
those for whom D-Lib Magazine had a significant career impact, both as
readers and authors; after the editors we were the top two most frequent
authors, with 39 unique contributions between us. \\

Internet-based digital libraries (or ``electronic libraries'' as
they were frequently known as prior to 1994) predated the popularity
of the web; some of the well-known examples include: ``Knowbots''
\cite{knowbots}, the CORE electronic journal project \cite{lesk1991core},
Netlib \cite{dongarra1987distribution}, xxx.lanl.gov \cite{187185},
Computer Science Technical Reports Project (CS-TR) \cite{cs-tr:kahn},
Wide-Area Technical Report Server (WATERS) \cite{maly1995wat},
and the Langley Technical Report Server (LTRS) \cite{ltrs:4567}.
However, the NSF-funded Digital Library Initiative (DLI, 1994--1998)
co-occurred with the rapidly increasing interest in the web, which
was accelerated by the late 1993 release of the NCSA Mosaic browser
\cite{andreessen1994ncsa}.  As a result, the story of the early web
parallels the story of digital libraries and D-Lib Magazine.  It is in
this context of the nascent web that D-Lib Magazine should be understood,
for it addressed a critical need in 1995.  From the editorial of the
first issue \cite{friedlander:1995:editorial}:

\begin{quote}
The magazine is itself an experiment in electronic publishing, which fulfills its communication function for the Digital Library Forum by testing the limits of writing in and for a wholly networked environment. We have no -- and propose no -- print analogue, and we will be most intrigued by substantive articles that take advantage of the power of hypermedia while retaining the strengths of traditional, print publishing.
\end{quote}

The first issue had 14 ``Clips and Pointers'' -- announcements,
deadline reminders, calls for participation, requests for proposals and
papers, and brief updates.  Although email lists served these functions
(and still continue to do so), this announcement and awareness function
of a magazine has largely been replaced by blogs, social media.  One no
longer expects to learn of calls for papers or requests for proposals
in a magazine, and event summaries are now easily discoverable via
search engines with far more precision than those of the mid-1990s
(e.g., Lycos \cite{mauldin1997lycos}).  For example, our conference report for the 2003
Joint Conference on Digital Libraries (JCDL) was published in D-Lib
Magazine \cite{nelson2003rta}, but JCDL 2020 is best reviewed in blogs
\cite{ws-dl:jcdl2020} or Twitter\footnote{\url{https://twitter.com/search?q=\%23JCDL2020&src=typed_query&f=live}}. \\

The first issue had three articles, then carried under the heading
of ``stories and briefings'', reflecting the early position of a
``magazine'' and not an online journal.  In fact, they were summaries
of existing conventional reports and publications:

\begin{enumerate}

\item ``Metadata: the foundations of resource description'' -- a summary
of the OCLC/NCSA Metadata Workshop \cite{oclc-metdataworkshop} that
produced the Dublin Core Metadata Element Set, which continues today as
the Dublin Core Metadata Initiative (dublincore.org).

\item ``An agent-based architecture for digital libraries'' -- a description
of the distributed agent architecture explored in the University of
Michigan Digital Library (UMDL) \cite{dlib:dli:um}; the University of Michigan was one of six
participants in the first NSF/DARPA/NASA Digital Library Initiative (DLI).

\item ``Key concepts in the architecture of the digital library'' -- an
introduction to and contextualization of what would become known as the
``Kahn-Wilensky Framework'' (KWF) \cite{kwf}, part of which included
handles \cite{lannom:handles}, upon which Digital Object Identifiers
\cite{paskin:doi:1999} are implemented.

\end{enumerate}

As tentative steps in this new publishing experiment, all three articles
are single authored (though they summarize multi-author publications),
are relatively short, and have limited figures and references.  Although
D-Lib Magazine would soon evolve into a venue where original research was
published (e.g., a 1999 editorial estimates that half of the contributions
described original research \cite{arms:1999:editorial}) and essentially
functioned as an online journal, it was edited and never refereed.
This produced a well-known problem: if you wanted your material to reach
a wide audience, it needed to be in D-Lib Magazine, but if you wanted
academic ``credit'', it needed to be in a conventional journal or
refereed conference proceedings.  In the time before Google, Google
Scholar, CiteSeer, Microsoft Academic et al., this was a binary choice.
Now it is possible for authors to gain the imprimatur of a quality journal
or conference proceedings, and at the same time leverage the permissive
attitude regarding pre-prints and e-prints of many publishers (e.g.,
ACM) to ensure that articles are discoverable and freely available. \\

\section{D-Lib Magazine as a publishing experiment}
\label{sec:publishing-experiment}

D-Lib Magazine was unique in many respects.  First, although it clearly
billed itself as a ``magazine'', it quickly became a venue where
original research was published.  Second, although it initially offered
additional services and categories, the real innovation came about because
it embraced HTML, and only HTML, as the publication medium.  HTML allowed
the articles themselves to take advantage of a rapidly evolving medium,
including links and multimedia in a way PDF-primary publications could
not.  Finally, with the vantage of 25 years, the decisions made in how
D-Lib Magazine would be structured and maintained compare favorably to
other Web-based publishing peers which began shortly after D-Lib Magazine. \\

\subsection{More than a magazine, even if not quite a journal}

Although early issues had unsuccessful experiments with HyperNews
\cite{hypernews:1996} for comments as well as a separate ``technology playpen'' /
``technology spotlight'' section \cite{wilson:2005:editorial}, these features
were eventually subsumed within the HTML publishing experiment itself,
and D-Lib Magazine's primary unit of currency became its articles.
From 1995 through 2017, D-Lib Magazine published 265 issues and 1062
articles (D-Lib Magazine actually defined and evolved many different
categories of contributions \cite{wilson:2008:editorial}, but we refer
to entries available from the title index as simply ``articles'').
The issues were published monthly through June, 2006 (with the July/August
issues published simultaneously as ``7/8â''), and it switched to
bimonthly publication from July/August 2006 through July/August 2017.
D-Lib Magazine was always ``a magazine, not a peer-reviewed journal'' and
aimed for ``articles that are 1,500 to 3,000 words in length and seldom
accept articles in excess of 5,000 words'' \cite{wilson:2005:editorial}.
To explore this, we took the title index:

\begin{enumerate}

\item from the HTML extracted all links that begin with \texttt{<p class="archive">}, which includes the articles but excludes ``in brief'' and ``opinion'' entries.

\item for each of the 1062 URLs, we used \texttt{lynx -dump \$URL > \$filename}, which saves only the result of rendering the HTML into plain text.

\item used \texttt{wc -w} on each of the resulting files to count the number of words in the article.

\end{enumerate}

\begin{figure} 
\begin{center}
\includegraphics[scale=0.45]{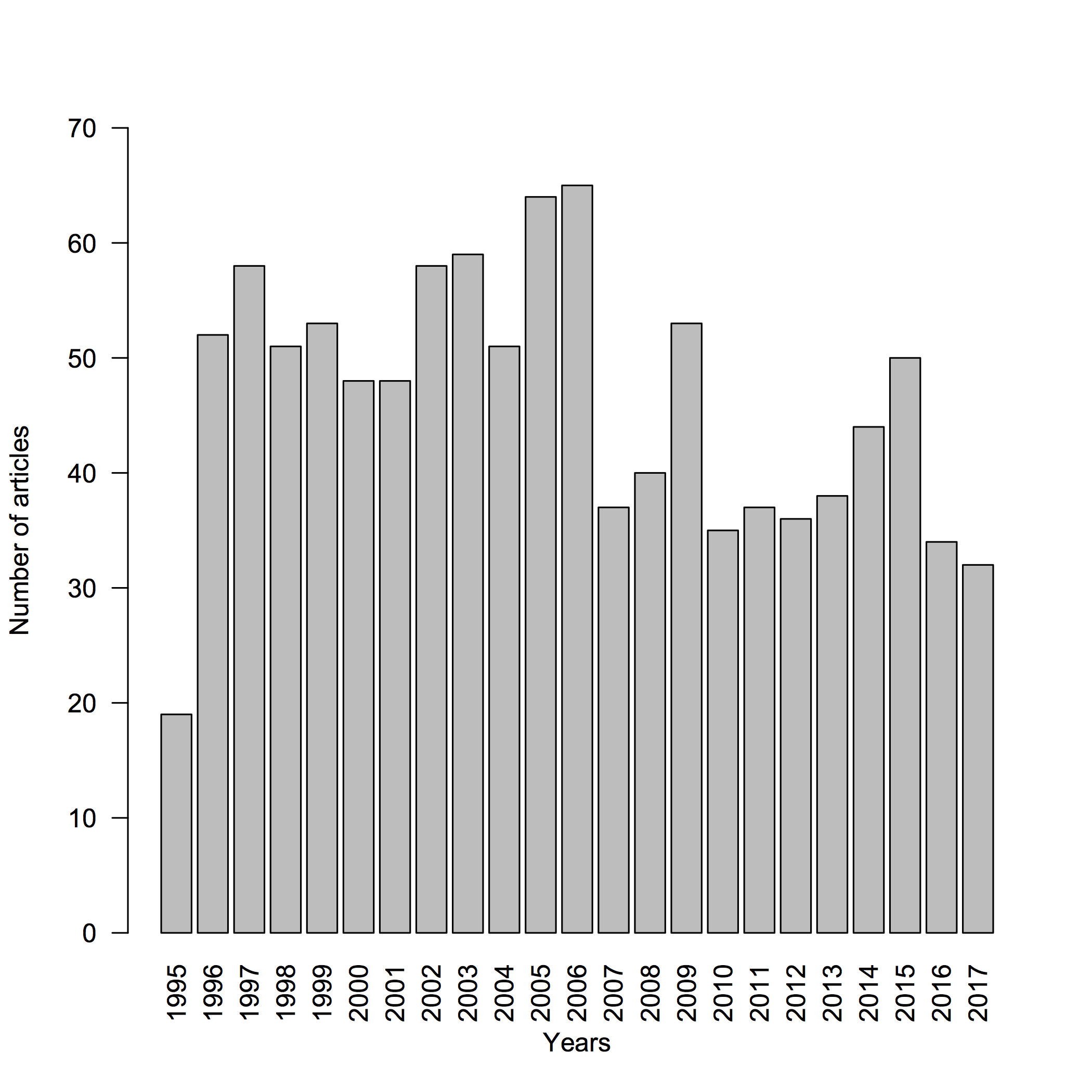}
\caption{Total articles published per year (1995 and 2017 were incomplete years).}
\label{fig:articles-per-year}
\end{center}
\end{figure}

\begin{figure} 
\begin{center}
\includegraphics[scale=0.45]{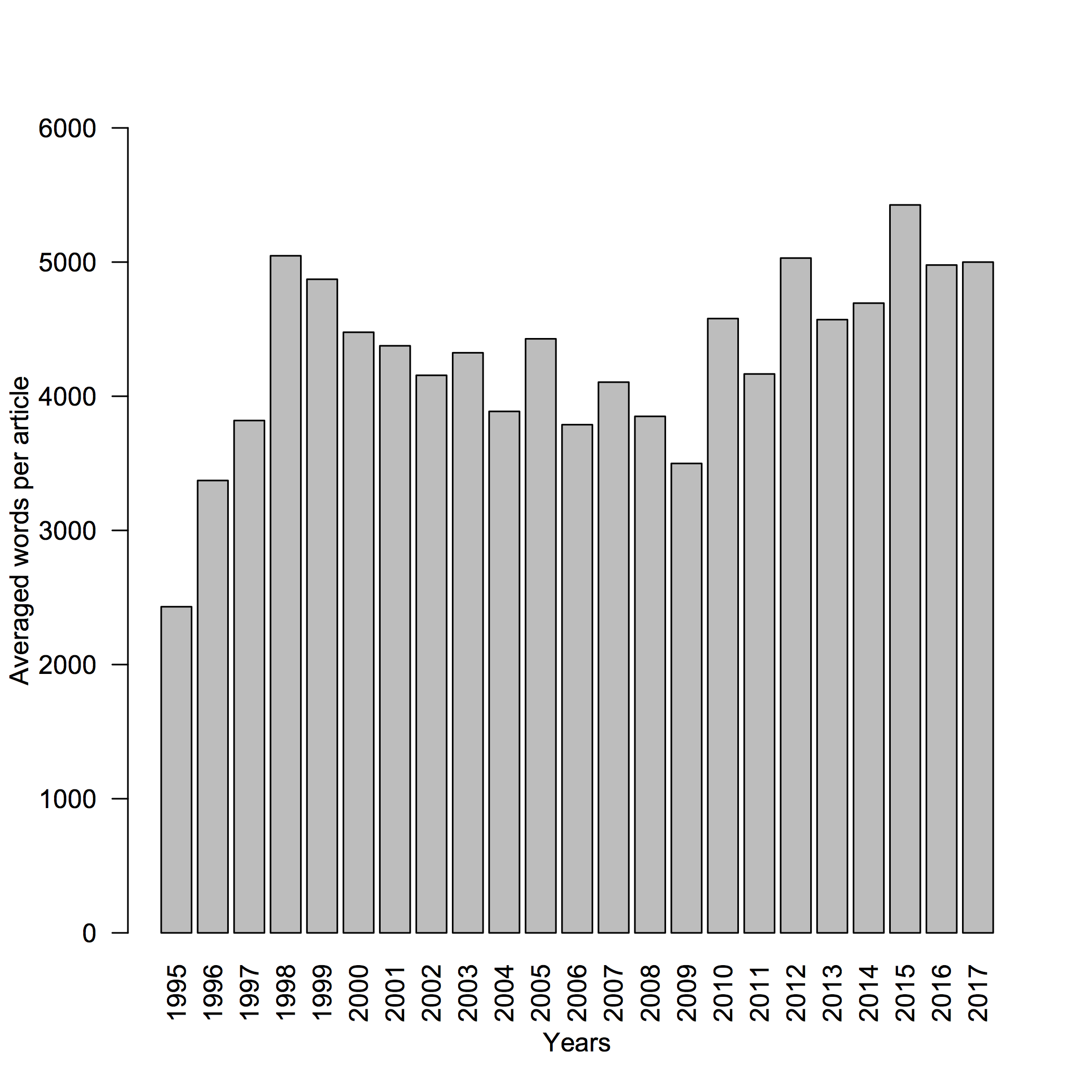}
\caption{Average words per article per year.}
\label{fig:words-per-year}
\end{center}
\end{figure}

Using lynx to render the HTML is not perfect, but it reasonably
approximates the number of words in the article.  Figure \ref{fig:articles-per-year} shows
the number of articles published each calendar year, and Figure \ref{fig:words-per-year}
shows the average number of words per article for each calendar year.
From Figure \ref{fig:words-per-year} we can see that although switching to bimonthly publication
in 2006 reduced the number of articles per year, it did not halve it.
Even though in 2017 D-Lib published only four issues (instead of six),
the total number was only slightly down from 2016, perhaps indicating
clearing the queue of remaining articles for the year. \\

Figure \ref{fig:words-per-year} shows a trend of shorter articles in
the first three years, and then finally hitting its stride in 1998,
perhaps corresponding with the acceptance of the format by both authors
and editors.  From 1998 on, the values fluctuate (we are unsure of why
2009 has a low value) but it is not until the last six years (2012--2017)
that the word count approximates the early peak from 1998.  \\

Even though it was never peer-reviewed, and did not have an editorial
board like a conventional journal (though it did have an advisory board\footnote{\url{http://web.archive.org/web/20000226003334/http://www.dlib.org/forum/advisory-board.html}}),
D-Lib Magazine had a significant impact in the conventional literature and
served as a de facto journal.  A ten year anniversary analysis (from 2005)
showed that D-Lib Magazine had acquired 147 citations from the ACM/IEEE
Joint Conference on Digital Libraries and its predecessor conferences
\cite{wilson:2005:editorial}.  A more detailed authorship and citation analysis
showed over 1300 citations in the first 15 years \cite{park:dlib:2010}.\\

\begin{figure}
\begin{center}
\includegraphics[scale=0.35]{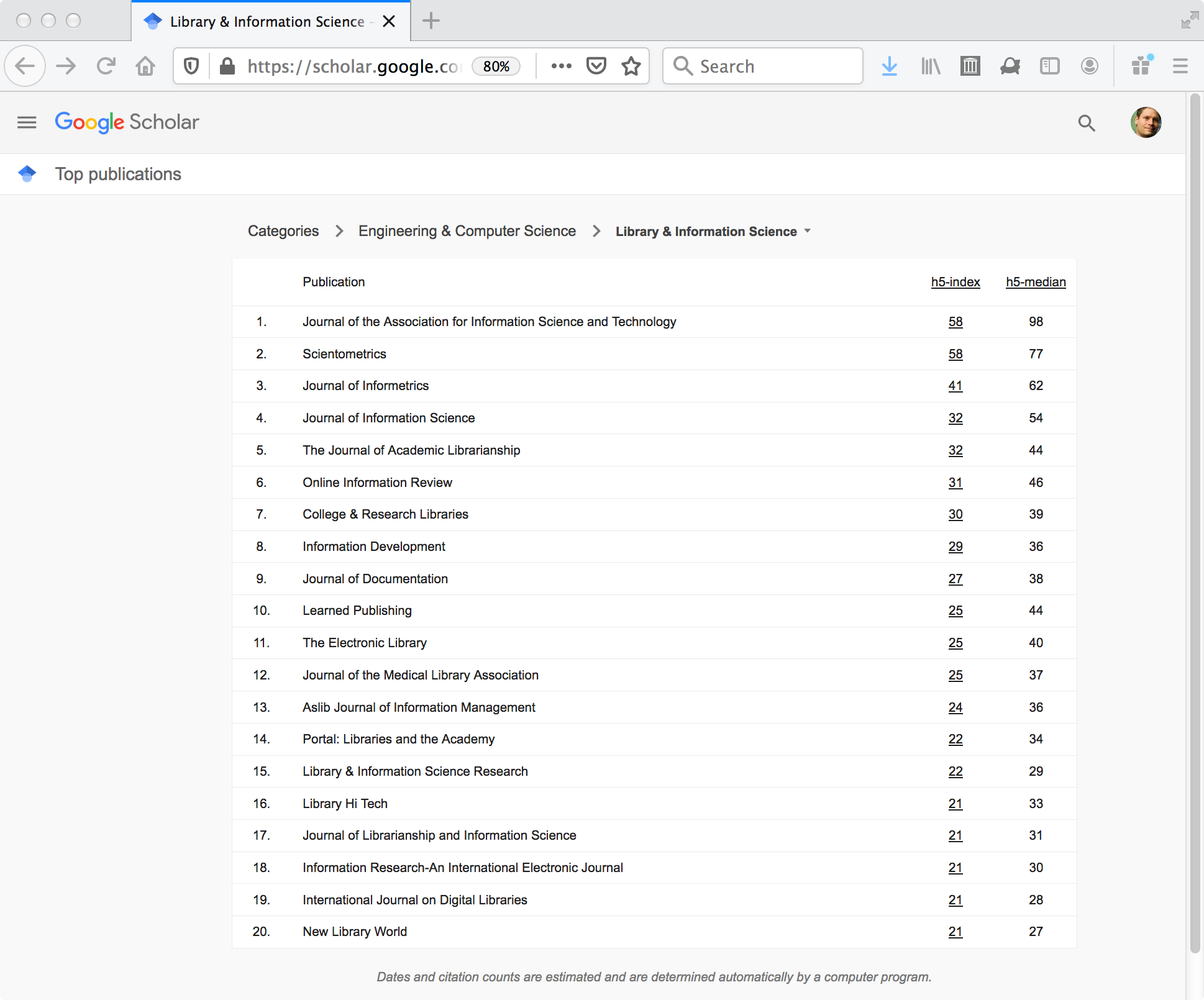}
\caption{Library \& Information Science, Google Scholar \url{https://scholar.google.com/citations?view\_op=top\_venues&hl=en&vq=eng\_libraryinformationscience} (from 2020-07-19).}
\label{fig:gs-lis-2020}
\end{center}
\end{figure}

\begin{figure}
\begin{center}
\includegraphics[scale=0.29]{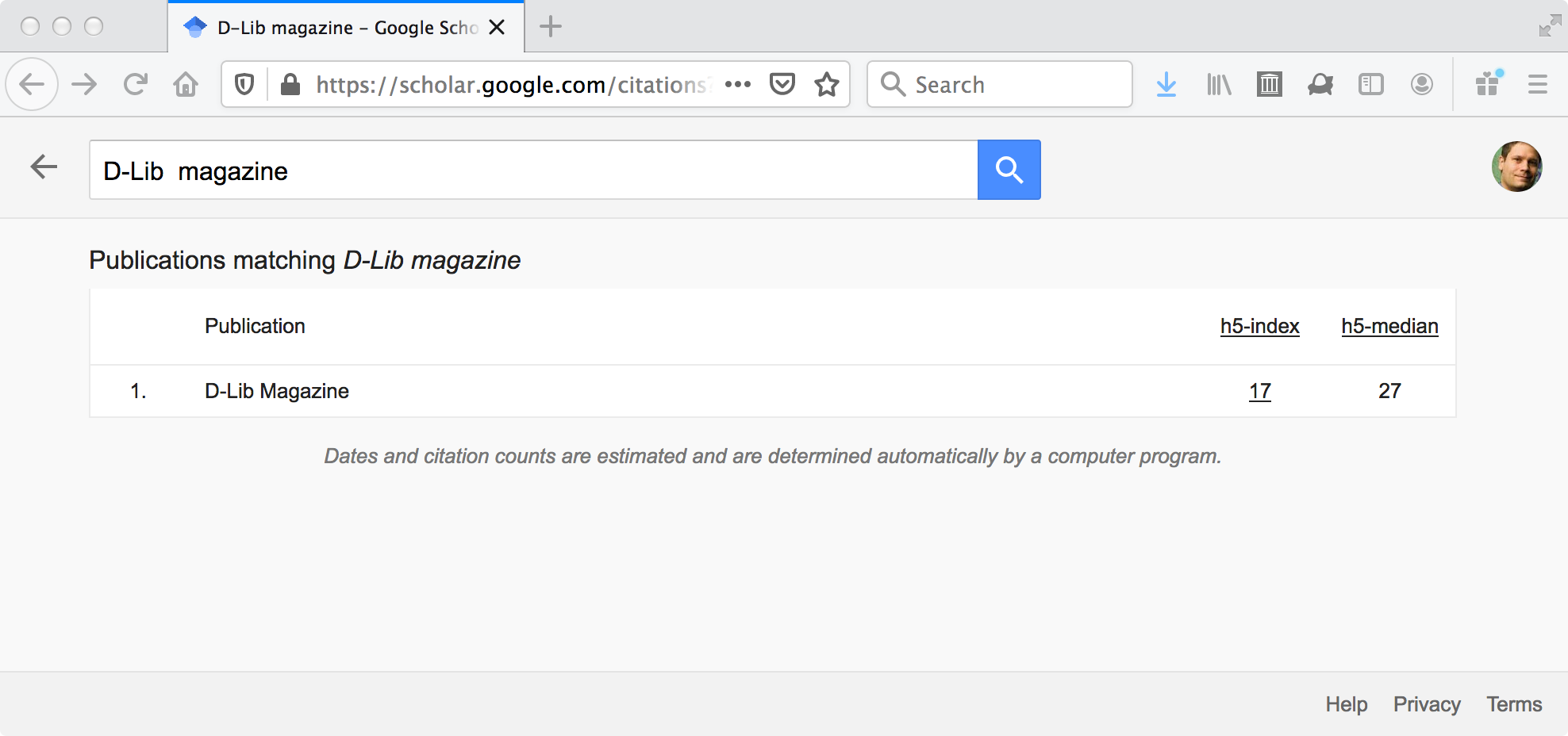}
\caption{D-Lib Magazine, Google Scholar \url{https://scholar.google.com/citations?hl=en&view_op=search_venues&vq=D-Lib++magazine&btnG=} (from 2020-07-19).}
\label{fig:gs-dlib-2020}
\end{center}
\end{figure}

\begin{figure}
\begin{center}
\includegraphics[scale=0.29]{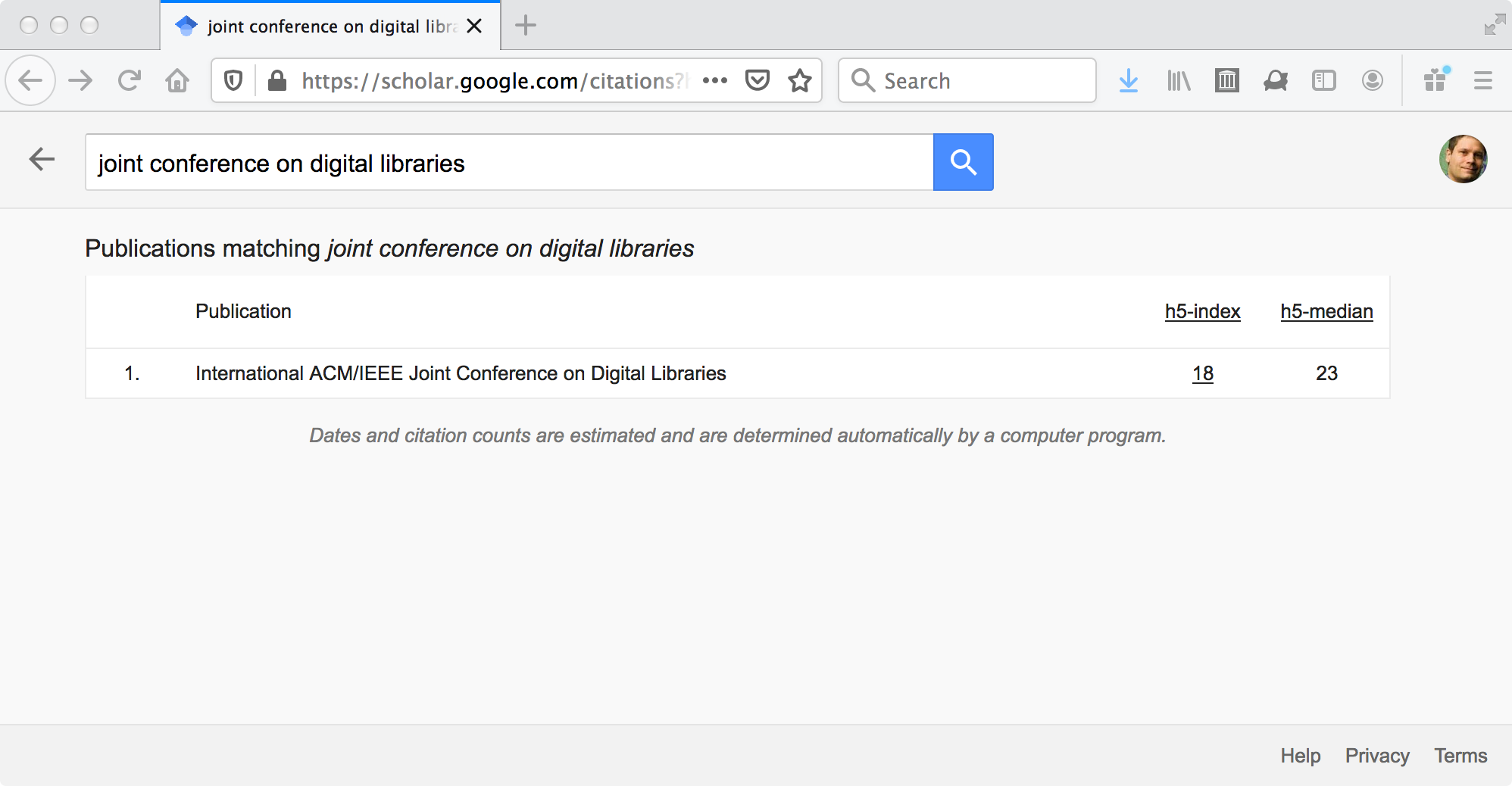}
\caption{JCDL, Google Scholar \url{https://scholar.google.com/citations?hl=en&view_op=search_venues&vq=joint+conference+on+digital+libraries&btnG=} (from 2020-07-19).}
\label{fig:gs-jcdl-2020}
\end{center}
\end{figure}

\begin{figure}
\begin{center}
\includegraphics[scale=0.29]{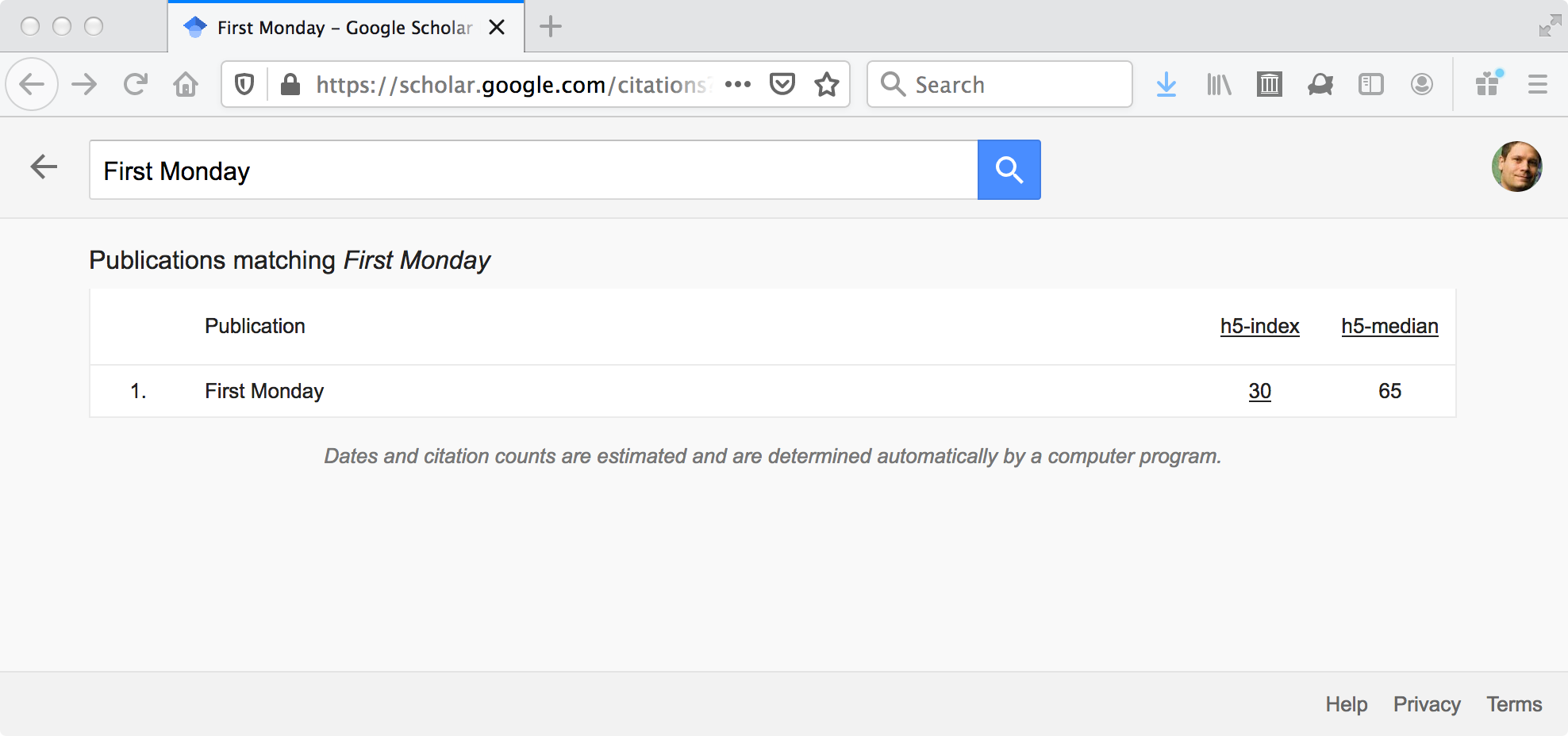}
\caption{First Monday, Google Scholar \url{https://scholar.google.com/citations?hl=en&view_op=search_venues&vq=First+Monday&btnG=} (from 2020-07-19).}
\label{fig:gs-firstmonday-2020}
\end{center}
\end{figure}

A look at the 2020 Google Scholar rankings in ``Library \& Information
Sciences'' shows the top 20 venues in the field (Figure \ref{fig:gs-lis-2020}; note: ``digital
libraries'' as a field awkwardly straddles ``Library \& Information
Sciences'' and ``Databases \& Information Systems'' in Google
Scholar's classification).  D-Lib Magazine is just outside the top 20,
despite no new publications since 2017, with an h5-index of 17 (Figure \ref{fig:gs-dlib-2020}), which
is comparable to JCDL's h5-index of 18 (Figure \ref{fig:gs-jcdl-2020}).  Among its contemporaries
(section \ref{sec:peers}), First Monday is doing well with an h5-index of 30 (Figure \ref{fig:gs-firstmonday-2020}), but
we cannot determine in which category Google Scholar places First Monday.
Ariadne and the Journal of Digital Information are not included in Google Scholar's 2020 rankings.  \\

\subsection{Innovations in Web-based publishing}
\label{sec:innovations}

As the initial editorial makes clear, D-Lib Magazine was an ongoing
experiment in ``electronic publishing'' itself, and as a result was an
early adopter and proof-of-concept for a lot of conventions and techniques
that are now best practices in the community.  Perhaps most importantly,
D-Lib Magazine was always published in HTML -- and only in HTML:
there was never a parallel PDF version.  Submissions were encouraged in MS
Word\footnote{\url{http://web.archive.org/web/20000613151426/http://www.dlib.org:80/dlib/author-guidelines.html}},
but the editors handled the conversion to HTML themselves.  Adopting an
HTML-only publishing strategy seems obvious in retrospect, but considering
the limitations of HTML ca. 1995 (cf. HTML5 \cite{html5:w3c} today) this
was a bold strategy.  Despite the dominance of the PDF in the scholarly
publishing ecosystem, the HTML format allowed authors to experiment
with multimedia and interactivity extensions not possible with PDFs.
Quoting from the October, 1995 editorial, one gets a glimpse of the
willingness to explore the boundaries of what an HTML-only publication
could be \cite{friedlander:1995:editorial:oct}:

\begin{quote}
You will see that the stories have varied in their treatment of images,
for example, in the background color, and even in the organization of
the text itself. But I do not believe that these individual treatments
posed a problem for our readers, partly because the stories are
unified by subject, partly because the medium is itself experimental
and preconceptions are fairly few, and partly because in each case,
the structure of the story reinforces and extends its informational
content. Thus, the highly visual story that the Informedia team wrote
on indexing video\footnote{\url{http://www.dlib.org/dlib/july96/07wactlar.html}} subtly embodies the notion of frames in its file
structure. It offers readers multiple paths through the material
and cues through buttons not unlike the signage found in museums and
airports, and through menus that other writers for the magazine have
also employed. In the same issue, the Netlib authors used a classic,
straightforward narrative approach with an internal menu to explain the
complex structure of a library of mathematical software\footnote{\url{http://www.dlib.org/dlib/september95/netlib/09browne.html}}.
\end{quote}

As authors, we certainly appreciated the editors' willingness to
explore what new features were possible in an HTML scholarly publication.
For example, in our 1999 article about the Universal Preprint Service
\cite{oaipmh:ups}, we included screen cams to show the now defunct ups.cs.odu.edu
digital library in action.  Those screen cams were stored in .exe
format and would thus likely require emulation to run now, but those
animations (stored at dlib.org) would not have been possible in a PDF.
Another of our articles from 2002 used animations, but this time in a
more web-friendly and standard MPEG format \cite{nelson:object}.  In a 2005 article, we
did not use animations, but did have 377 images linked from the article,
a feat that would have been unwieldy at best in PDF \cite{bollen2005tad}.  Our last
article in D-Lib Magazine used JavaScript to make annotated hyperlinks
in the article actionable, thereby serving as a demonstration of how
``Robust Links''\footnote{\url{https://robustlinks.mementoweb.org}} could work in practice \cite{dlib:2015:reminiscing}.  \\

\begin{figure}
\begin{center}
\includegraphics[scale=0.25]{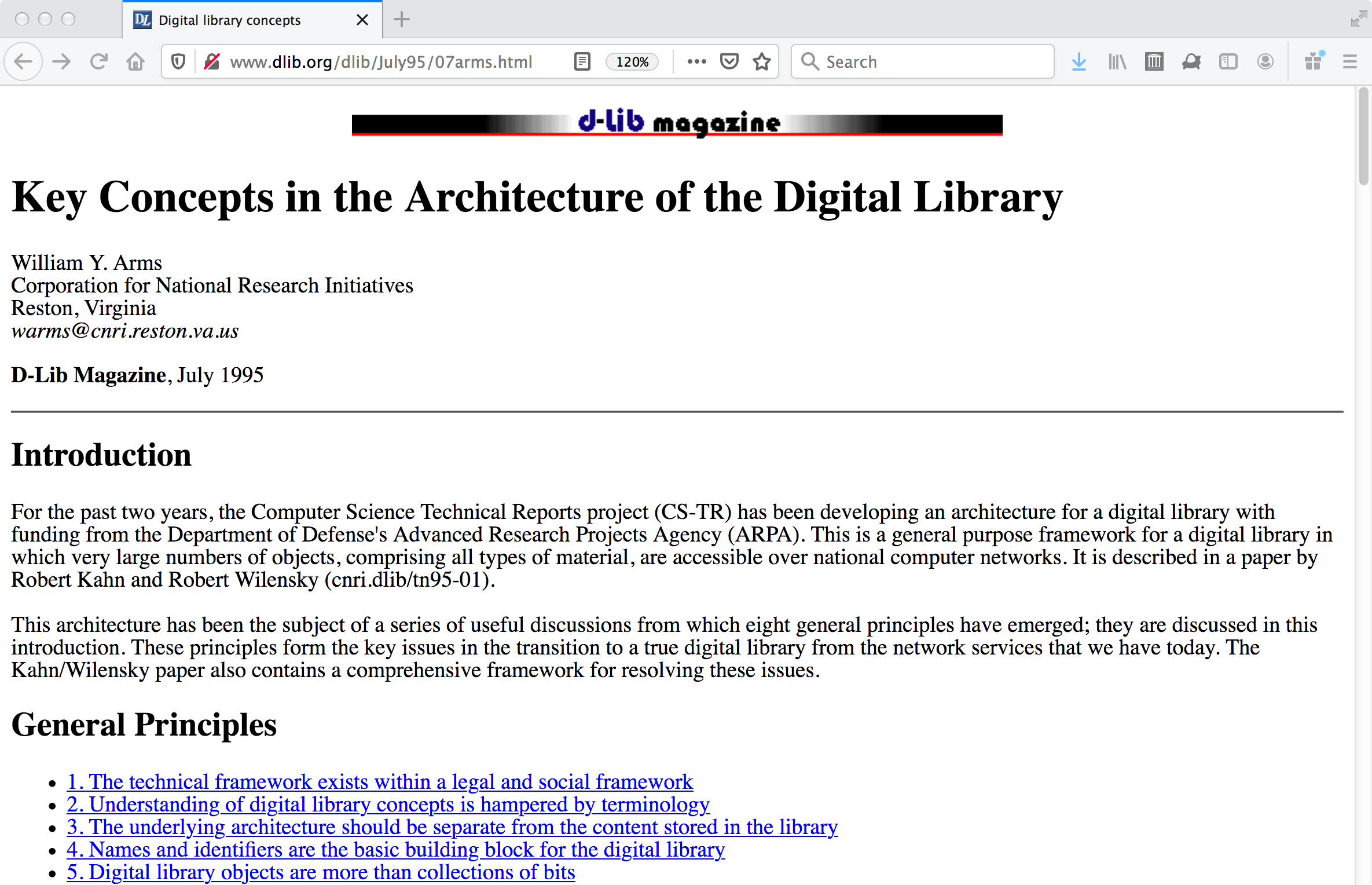}
\caption{D-Lib Magazine, live web: \url{http://www.dlib.org/dlib/July95/07arms.html}.}
\label{fig:dlib-live-web}
\end{center}
\end{figure}

\begin{figure}
\begin{center}
\includegraphics[scale=0.25]{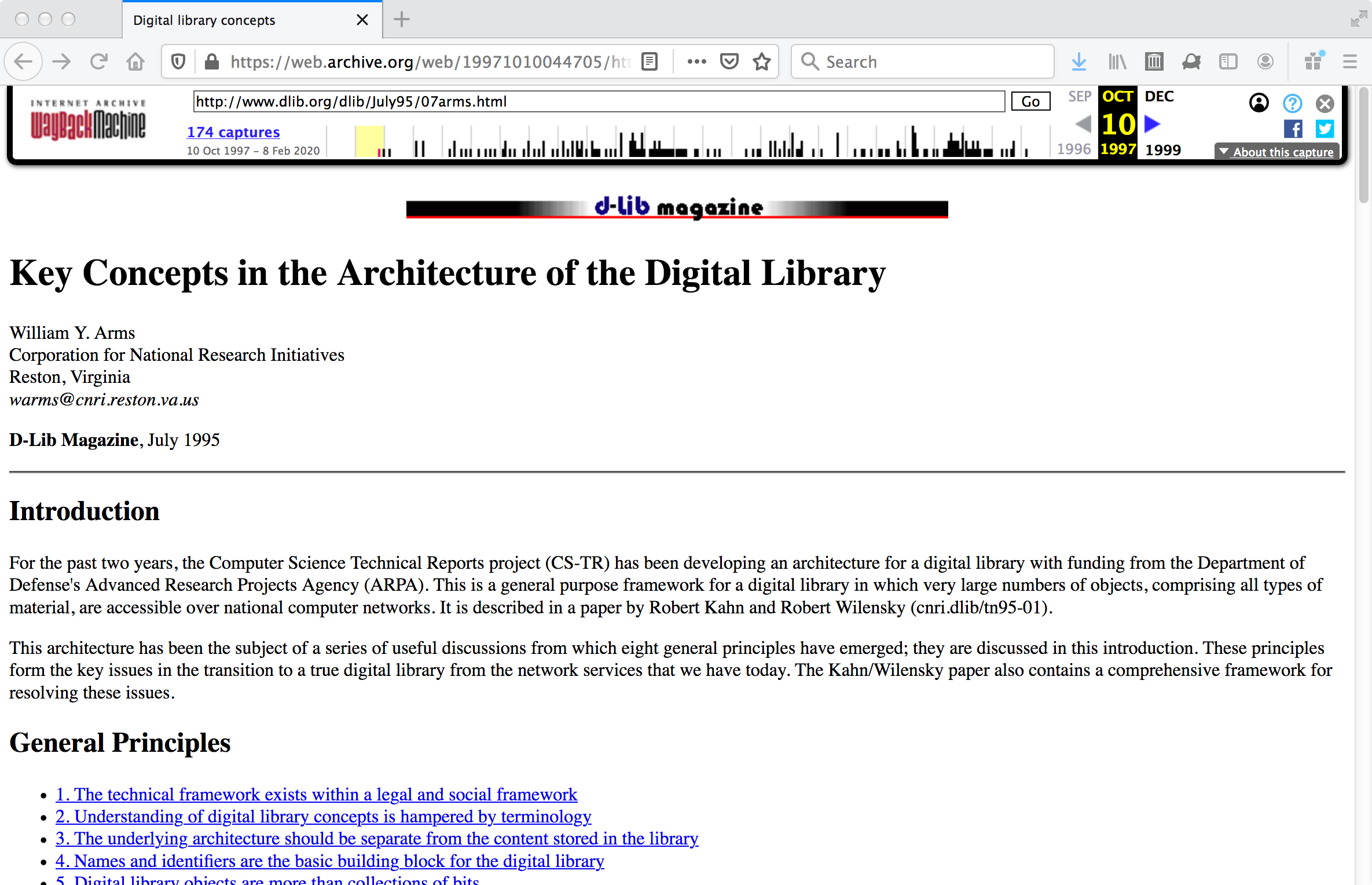}
\caption{D-Lib Magazine, archived in 1997: \url{https://web.archive.org/web/19971010044705/http://www.dlib.org/dlib/July95/07arms.html}.}
\label{fig:dlib-archived}
\end{center}
\end{figure}

Another significant decision was to fix the template and
formatting of past issues, and not reformat earlier issues
with updated templates.  Updates were only made in the cases of
errata and corrigenda\footnote{Although we thought we remembered
this policy being explicitly stated somewhere, we could find no
record of it.  In emails with former editors Larry Lannom and
Cathy Rey, neither could recall such a document.  The closest
we could find was ``Once the issue has been released, only vital
corrections or changes will be made to the file. These changes will
be noted and dated at the end of the file.'' in the Author Guidelines:
\url{http://web.archive.org/web/20000613151426/http://www.dlib.org/dlib/author-guidelines.html}.}.
D-Lib Magazine updated their design as tools and experience allowed, but
the first issue looks the same today as it did 25 years ago, thereby
serving as a monument to the best practices of the time.  Indeed,
the live web version of the first issue and the web archived version of
the first issue are indistinguishable (Figures \ref{fig:dlib-live-web},
\ref{fig:dlib-archived}, \ref{fig:dlib-curl-md5}).  Not only did they keep their
HTML and style intact, but thanks to an ongoing
commitment from CNRI all of D-Lib Magazine's issues are still available on
the live web, with no changes in their URIs since the fourth issue (October, 1995)\footnote{The first three issues were published at \url{http://www.cnri.reston.va.us/home/dlib.html} (cf. \url{https://www3.wcl.american.edu/cni/9507/6207.html}), and it was not until the October, 1995 issue that \url{www.dlib.org} was adopted (``Please note that D-Lib has a new address: http://www.dlib.org'' -- \url{http://www.dlib.org/dlib/october95/10contents.html}).}.
Although we have long known ``Cool
URIs Don't Change'' \cite{berners-lee:cool},
the reality is that most do, and persisting over 5,000
URIs\footnote{\url{https://www.google.com/search?q=site:dlib.org}}
for up to 25 years is an accomplishment in itself.  \\

\begin{figure}
\begin{scriptsize}
\begin{verbatim}
% date
Mon Jul 13 12:44:38 EDT 2020
% curl -s http://www.dlib.org/dlib/July95/07arms.html | md5sum
3cc0fb32a7fe8f1f4de9a40aa5069cfe  -
% curl -s https://web.archive.org/web/19971010044705id_/http://www.dlib.org/dlib/July95/07arms.html 
| md5sum
3cc0fb32a7fe8f1f4de9a40aa5069cfe  -
\end{verbatim}
\end{scriptsize}
\caption{Using curl to download both the live web version and first archived version (from 1997 and in ``raw'' format, via \texttt{id\_}) and show they produce the same md5 hash.}
\label{fig:dlib-curl-md5}
\end{figure}

Another groundbreaking innovation for D-Lib Magazine was that it was open
access before that term was even coined, with the authors retaining their
copyright, and D-Lib requiring neither subscriptions for readers nor
article processing charges from the authors.  This ensured it reached
a wide audience, both authors and readers, but it also resulted in
chronic funding problems after the expiration of the initial grants
that supported the D-Lib Forum ended.  In an editorial for the ten year
anniversary issue \cite{kahn:2005:editorial}, Robert Kahn said:

\begin{quote}
Producing a high quality magazine on the net each month turned out to be
somewhat less difficult than I would have expected, due almost entirely to
the quality of the editorial staff and the willingness of the readership
to contribute interesting articles. Funding the continued production of
the magazine has been, perhaps, its biggest challenge. While the initial
funding from DARPA covered most of the early costs, DARPA was unable to
continue the support indefinitely. Subsequent funding from NSF helped
greatly, but covered perhaps half the ongoing costs, with CNRI picking
up the other half.
\end{quote}

Although subscriptions and author fees were considered \cite{lannom:2006:editorial}, they
were never implemented. In 2007, the ``D-Lib Alliance'' membership
organization was created \cite{wilson:2007:editorial} that assisted with funding, but the
final issue in July 2017 acknowledged that decreased financial support
was part of the reason for ceasing publication \cite{lannom:2017:editorial}:

\begin{quote}
Financial support for the magazine has waned over recent years, the number of unsolicited high quality articles thrown over our transom has declined, and the very phrase 'Digital Libraries' has gone from sounding innovative to sounding a bit redundant. In short, it seemed like time to make a graceful exit.
\end{quote}

Another innovation that resulted from open access HTML-only publishing
was D-Lib being the first venue to have its handles (and later DOIs,
to be discussed further in section \ref{sec:kwf}) resolve to articles
themselves, not a landing page describing the article.  By eschewing PDF,
the format of paywalls, D-Lib Magazine was able to subtly reinforce
that its content was part of the Web, and not something separate, to
be downloaded via the Web.  The ability to link and provide embedded
multimedia enables the scholarly object to enjoy the same advances (and
risks, such as link rot
\cite{jones2016scholarly,klein2014scholarly,mccown:availability}) as the
rest of the web.  Another subtle result of embracing handles (and DOIs)
is that although D-Lib Magazine was published as a conventional serial,
it also embraced persistent identifiers for individual articles (owing
from the computer science technical report heritage of CNRI's technology),
which facilitates the disaggregation of serials into articles that are
directly and persistently identifiable, which reinforces them as being
``on the Web'' as first-class citizens. \\

Another innovation D-Lib Magazine embraced was the use of site mirrors
allowing users in Europe and Asia to interact with geographically
closer mirrors for faster response.  That approach to address bandwidth
limitations was common at the time and is now solved via content delivery
networks (CDNs). Three of the D-Lib mirrors are still functioning, down
from a peak of five\footnote{\url{http://web.archive.org/web/20150224045836/mirror.dlib.org/about.html}}.  In addition to the utility the mirrors provide,
they were also presumably intended as demonstrators for more advanced
Handle resolution techniques, such as being able to resolve to one of
multiple URLs \cite{doihandbook}.  \\

\subsection{Other contemporary Web-based journals and magazines}
\label{sec:peers}

\begin{figure}
\begin{center}
\includegraphics[scale=0.25]{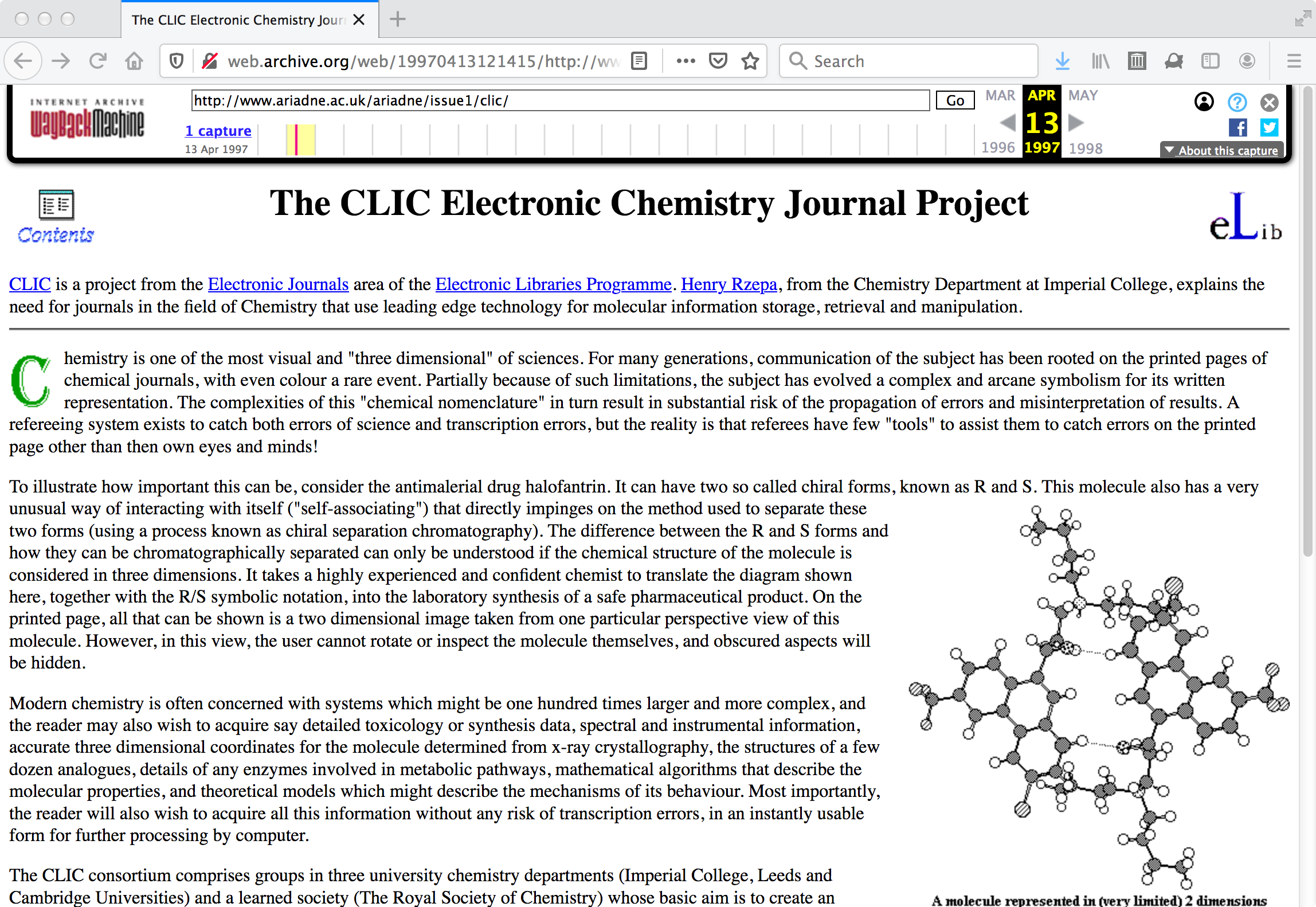}
\caption{An article from the first issue of Ariadne (published in 1996, archived in 1997 (\url{http://web.archive.org/web/19970413121415/http://www.ariadne.ac.uk/ariadne/issue1/clic/))}.}
\label{fig:ariadne-archived}
\end{center}
\end{figure}

\begin{figure}
\begin{center}
\includegraphics[scale=0.25]{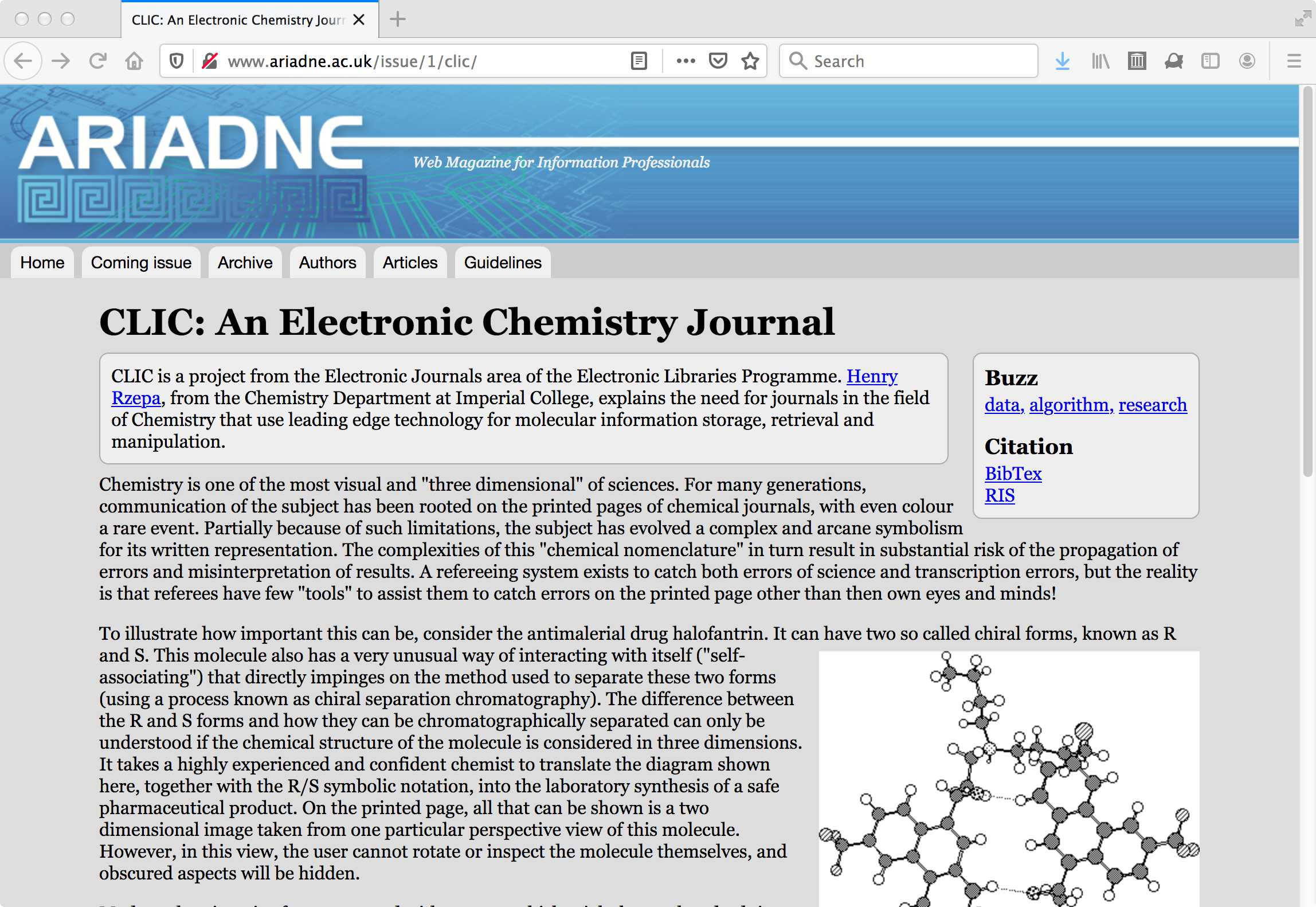}
\caption{The same article in 2020 (\url{http://www.ariadne.ac.uk/issue/1/clic/)}.}
\label{fig:ariadne-live}
\end{center}
\end{figure}

There were other contemporary experiments in on-line publishing
from generally the same community as well. For example,
Ariadne\footnote{\url{http://www.ariadne.ac.uk/}} is an online magazine
that began publishing in 1996 and is still publishing (78 issues since
1996).  It was similarly not peer-reviewed, aimed at practitioners,
and was initially funded by the Joint Information Systems Committee
(JISC, since renamed Jisc), a UK activity that can be considered roughly
analogous to the USA DLI program.  Ariadne also had an HTML focus from the
very beginning.  It has changed publishers a few times, as well as changed
its URIs and template through time (Figures \ref{fig:ariadne-archived},
\ref{fig:ariadne-live}, \ref{fig:ariadne-curl-404}).  It does not use handles or DOIs. \\

\begin{figure}
\begin{scriptsize}
\begin{verbatim}
$ curl -I http://www.ariadne.ac.uk/ariadne/issue1/clic/
HTTP/1.1 404 Not Found
Date: Sun, 19 Jul 2020 22:29:16 GMT
Server: Apache/2.4.6 (CentOS) OpenSSL/1.0.2k-fips PHP/7.2.24
Content-Type: text/html; charset=iso-8859-1
\end{verbatim}
\end{scriptsize}
\caption{The original URI for the first issue of Ariadne is 404.}
\label{fig:ariadne-curl-404}
\end{figure}

First Monday began in 1996 as a monthly peer-reviewed journal, and is
still being published.  But over time, its URIs have changed (from
firstmonday.dk to simultaneously firstmonday.org and a path within
journals.uic.edu), and its template changed along the way.  It uses DOIs,
and we believe it adopted them in 2013. \\

\begin{figure}\begin{center}
\includegraphics[scale=0.25]{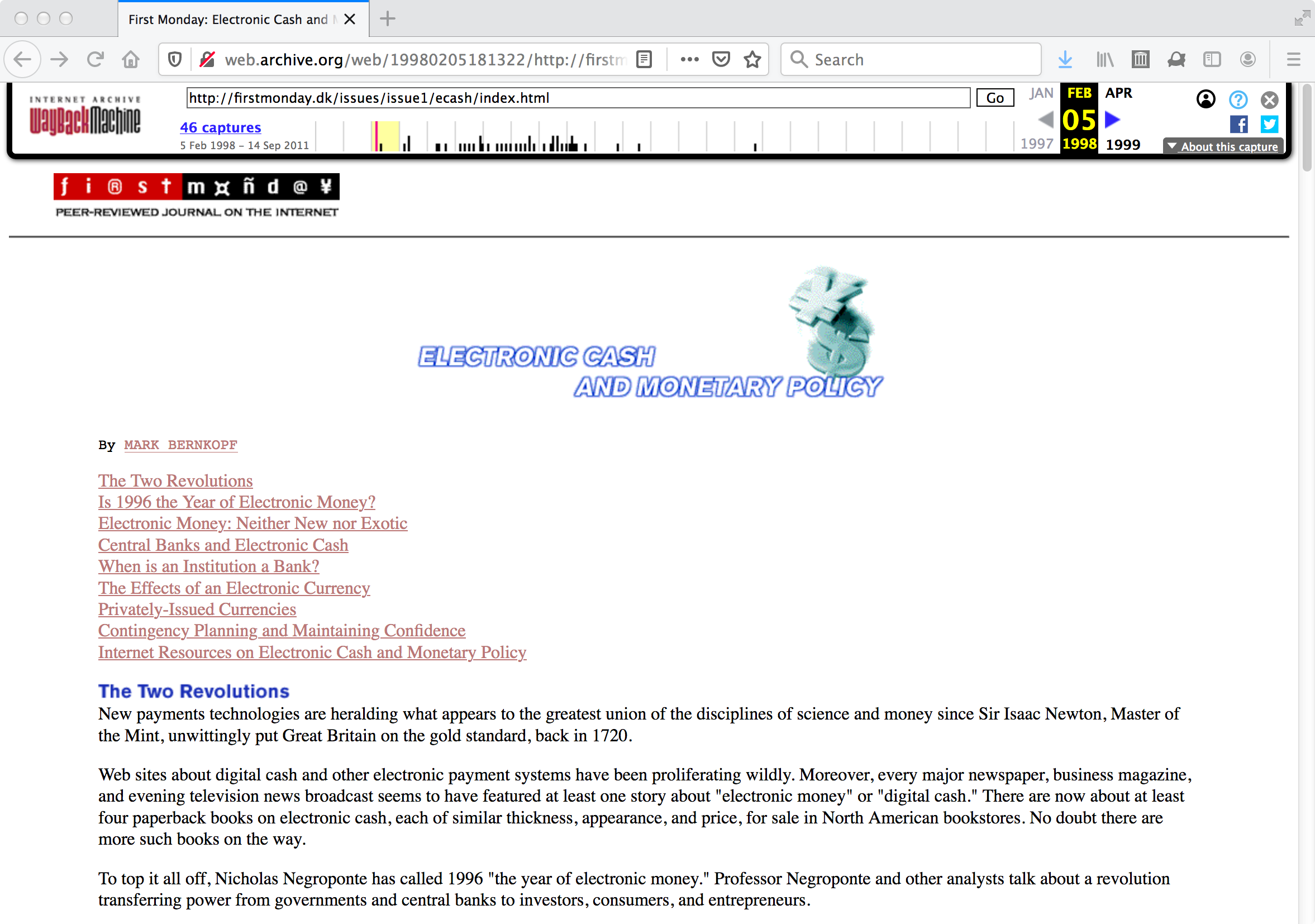}
\caption{An article from the first issue of First Monday, archived in 1998 (\url{web.archive.org/web/19980205181322/http://firstmonday.dk/issues/issue1/ecash/index.html)}.}
\label{fig:firstmonday-archived}
\end{center}
\end{figure}

\begin{figure}\begin{center}
\includegraphics[scale=0.25]{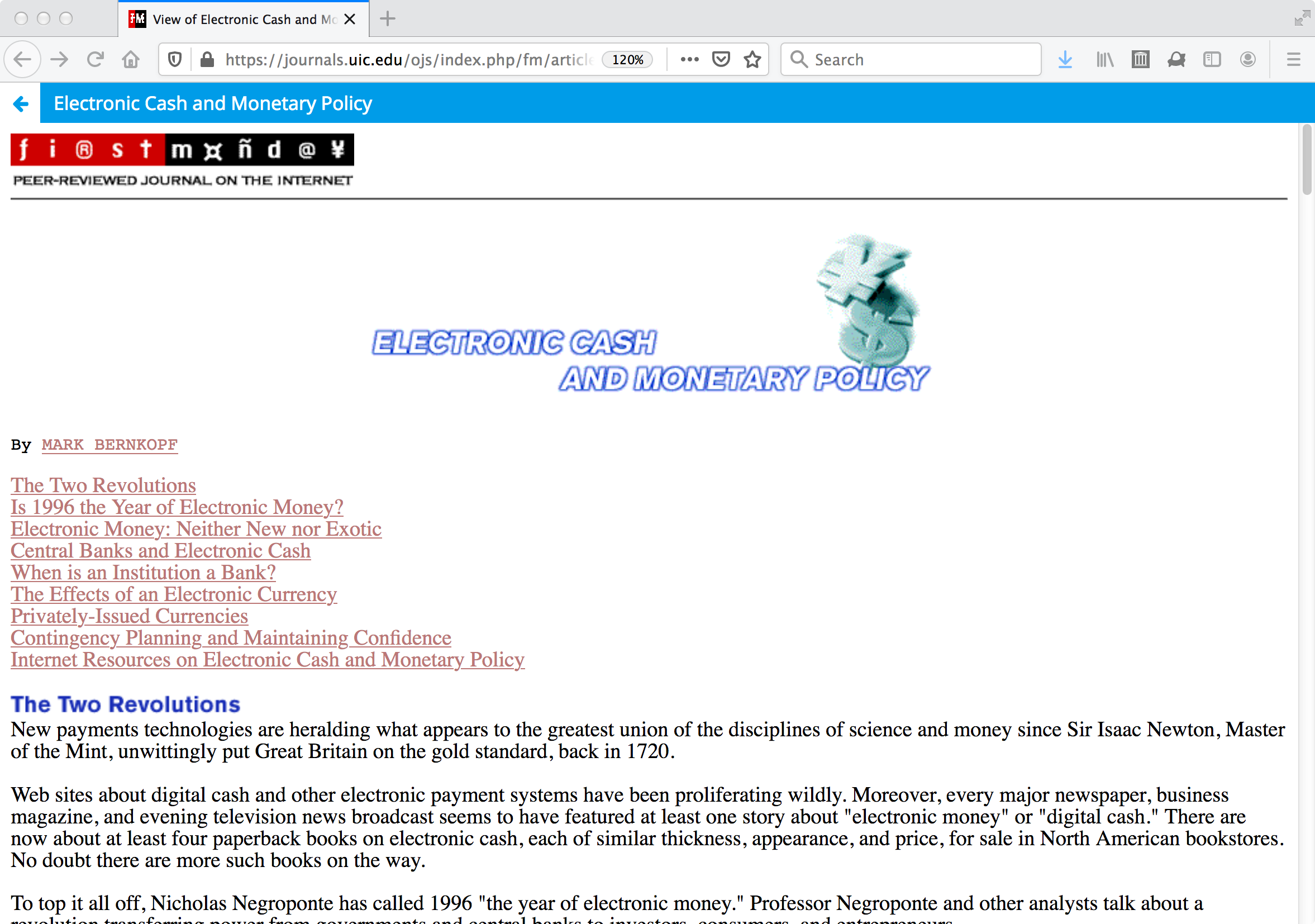}
\caption{A live Web version of the same article: \url{https://journals.uic.edu/ojs/index.php/fm/article/view/465/386}.}
\label{fig:firstmonday-live}
\end{center}
\end{figure}

\begin{figure}\begin{center}
\includegraphics[scale=0.25]{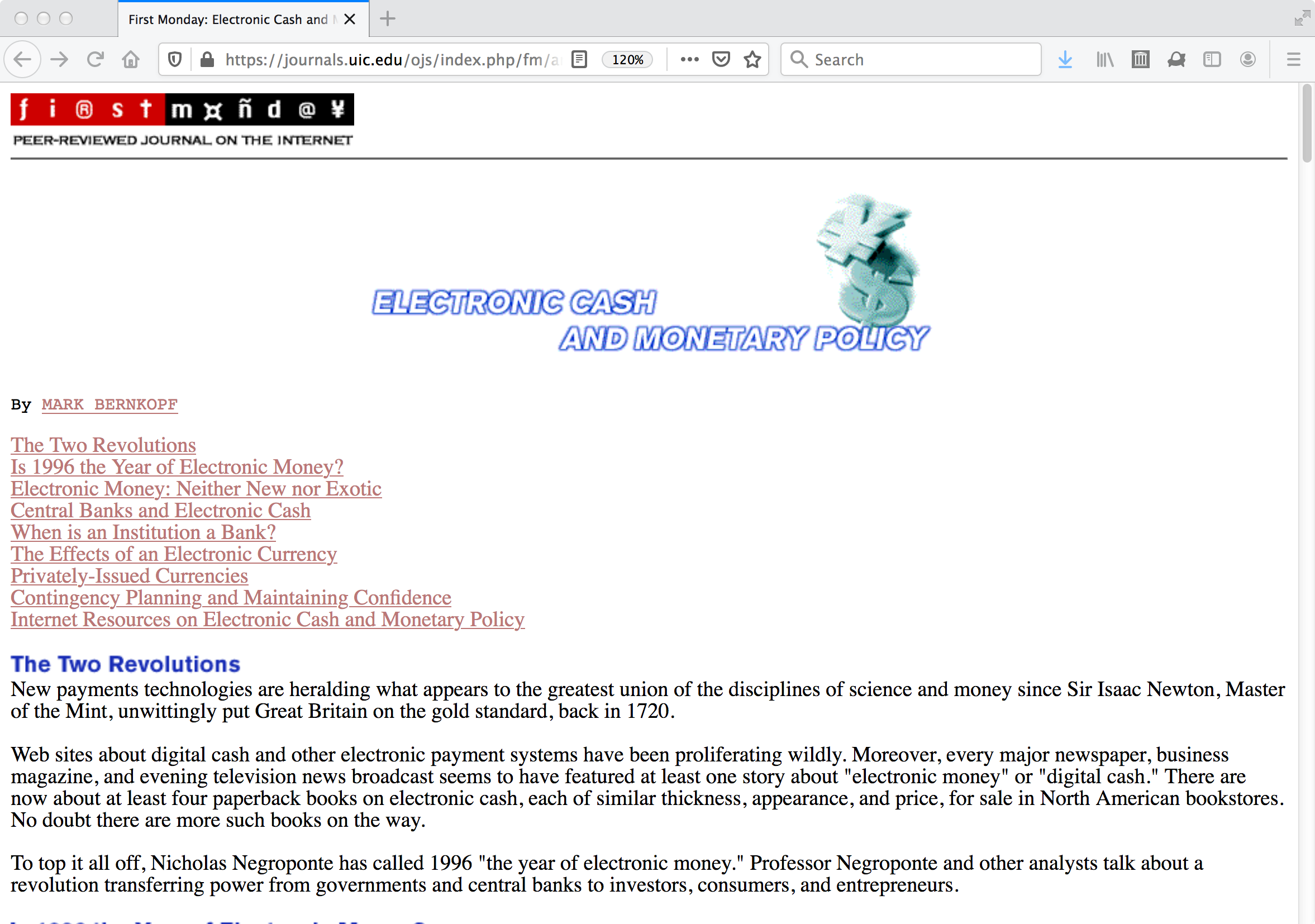}
\caption{The inner frame of the live Web version: \url{https://journals.uic.edu/ojs/index.php/fm/article/download/465/386?inline=1}.}
\label{fig:firstmonday-live-frame}
\end{center}
\end{figure}

There is more difference in the original First Monday (archived in 1998, Figure \ref{fig:firstmonday-archived}), the current live Web First Monday (Figure \ref{fig:firstmonday-live}), and the inner frame of the live Web First Monday (Figure \ref{fig:firstmonday-live-frame}) than first
appears, a result of significant reformatting of the articles over time.
Figure \ref{fig:firstmonday-wc} shows downloading the archived raw version (via \texttt{id\_}),
the live version, and the inner frame of the live version, respectively.
The Unix utility \texttt{wc} (word count) respectively shows the lines,
words, and characters of each file, all of which are significantly
different.  \\

\begin{figure}
\begin{scriptsize}
\begin{verbatim}
$ curl -sL web.archive.org/web/19980205181322id_/http://firstmonday.dk/issues/issue1/ecash/index.html 
> first-monday-old
$ curl -ksL https://journals.uic.edu/ojs/index.php/fm/article/view/465/386  > first-monday-now
$ curl -ksL https://journals.uic.edu/ojs/index.php/fm/article/download/465/386?inline=1 
> first-monday-now-frame
$ wc first-monday-*
      39     100    1941 first-monday-now
       1    4521   34585 first-monday-now-frame
       0    4379   32308 first-monday-old
      40    9000   68834 total
\end{verbatim}
\end{scriptsize}
\caption{The word count (wc) utility shows the differences in lines, words, and characters (respectively) for each version of the same article.}
\label{fig:firstmonday-wc}
\end{figure}

The Journal of Digital Information (JoDI) began as a peer-reviewed journal
in 1997, and ceased publication in 2013 after irregular publication of
46 issues.  While it was active it transitioned from the University of
Southampton (jodi.ecs.soton.ac.uk and journals.ecs.soton.ac.uk (the former
no longer resolves (Figure \ref{fig:jodi-curl})) to Texas A\&M University
and Texas Digital Library (journals.tdl.org).  The templates changed
through time, and publication was always a hybrid of either HTML or PDF. It did 
not use handles or DOIs. \\

\begin{figure}
\begin{scriptsize}
\begin{verbatim}
% curl -I http://jodi.ecs.soton.ac.uk/
curl: (6) Could not resolve host: jodi.ecs.soton.ac.uk
\end{verbatim}
\end{scriptsize}
\caption{\url{jodi.ecs.soton.ac.uk} no longer resolves.}
\label{fig:jodi-curl}
\end{figure}

\begin{figure}\begin{center}
\includegraphics[scale=0.25]{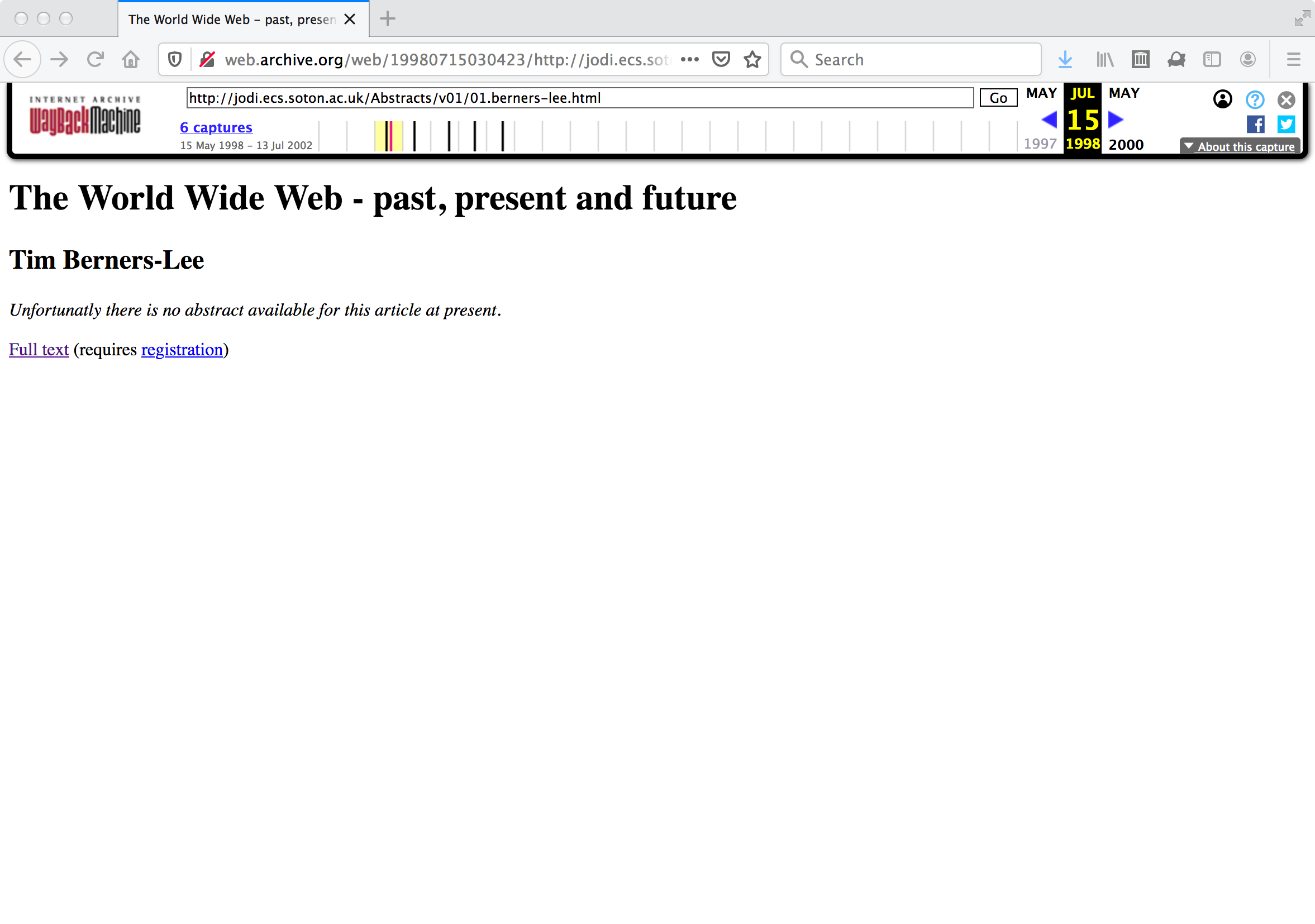}
\caption{A landing page for an article from issue 1 (1997), archived in 1998: \url{http://web.archive.org/web/19980715030423/http://jodi.ecs.soton.ac.uk/Abstracts/v01/01.berners-lee.html}.}
\label{fig:jodi-landing-archived}
\end{center}
\end{figure}

\begin{figure}\begin{center}
\includegraphics[scale=0.25]{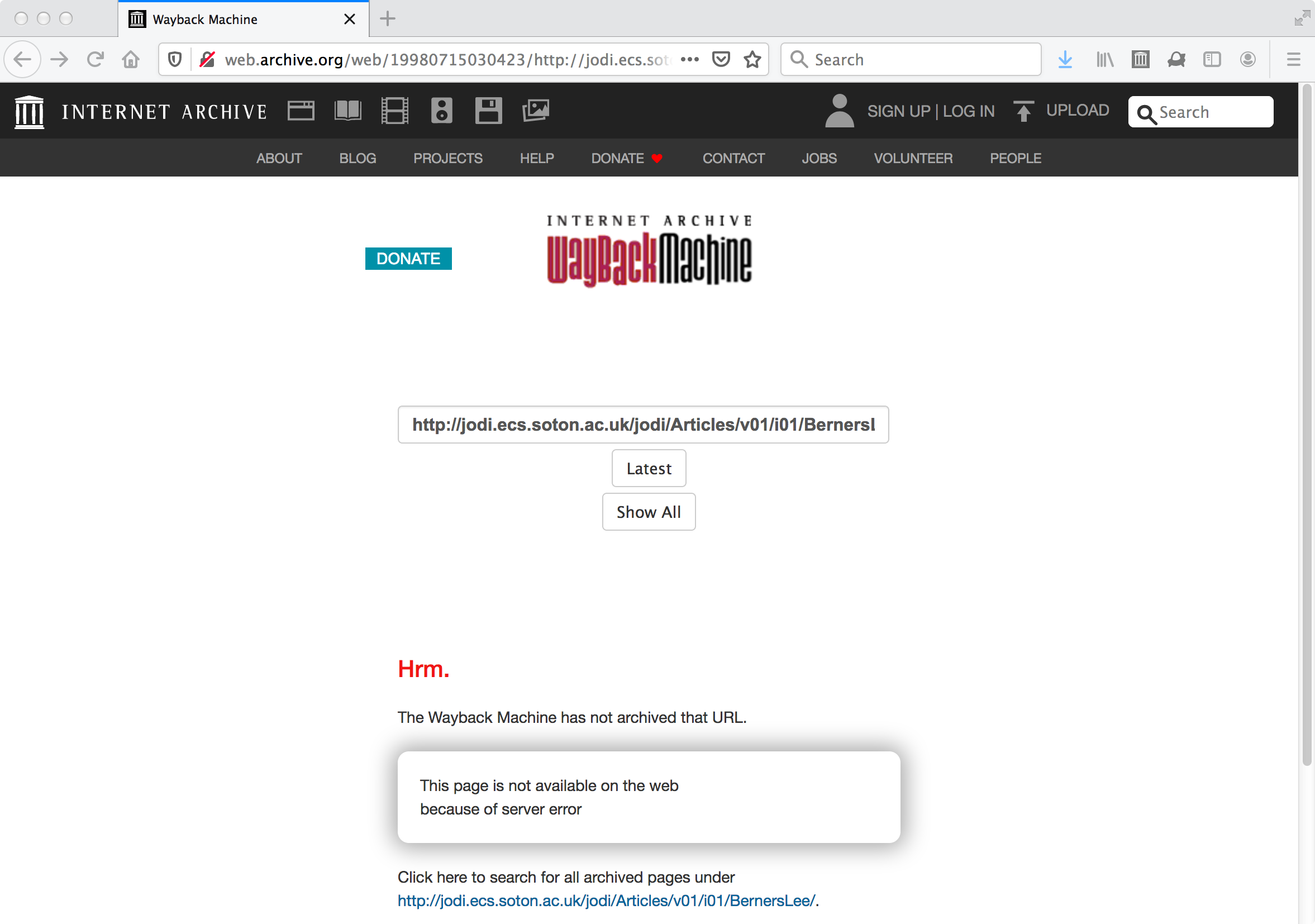}
\caption{Clicking through to: \url{http://web.archive.org/web/19980715030423/http://jodi.ecs.soton.ac.uk/jodi/Articles/v01/i01/BernersLee/} produces a 404 since this page was not on the surface web in 1998, and since the server jodi.ecs.soton.ac.uk is no longer on the live web, we cannot patch the archive.}
\label{fig:jodi-not-archived}
\end{center}
\end{figure}

\begin{figure}\begin{center}
\includegraphics[scale=0.25]{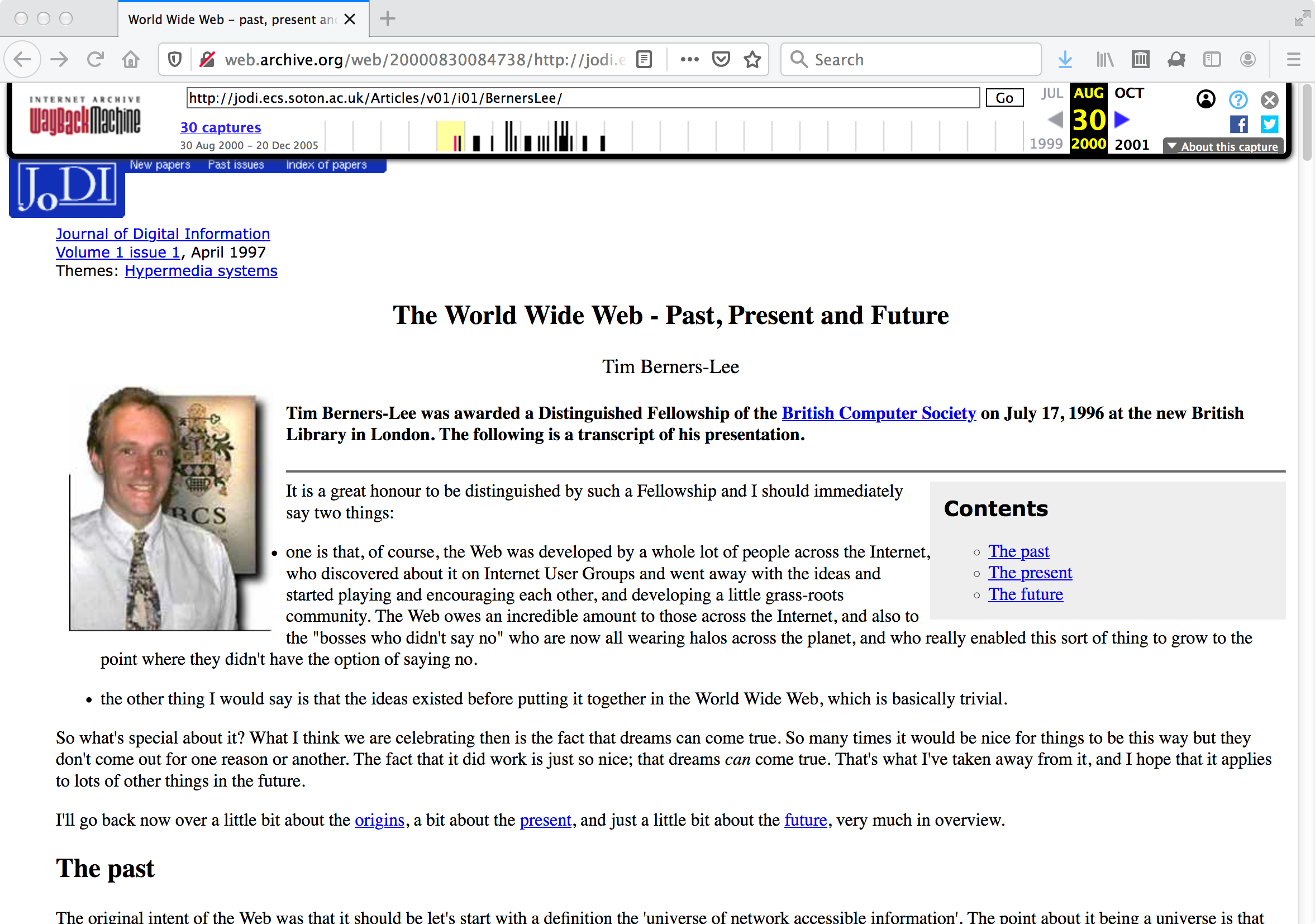}
\caption{In 2000 JoDI changed the URIs and removed login restrictions. \url{http://web.archive.org/web/20000830084738/http://jodi.ecs.soton.ac.uk/Articles/v01/i01/BernersLee/}.}
\label{fig:jodi-issue-archived}
\end{center}
\end{figure}

\begin{figure}\begin{center}
\includegraphics[scale=0.25]{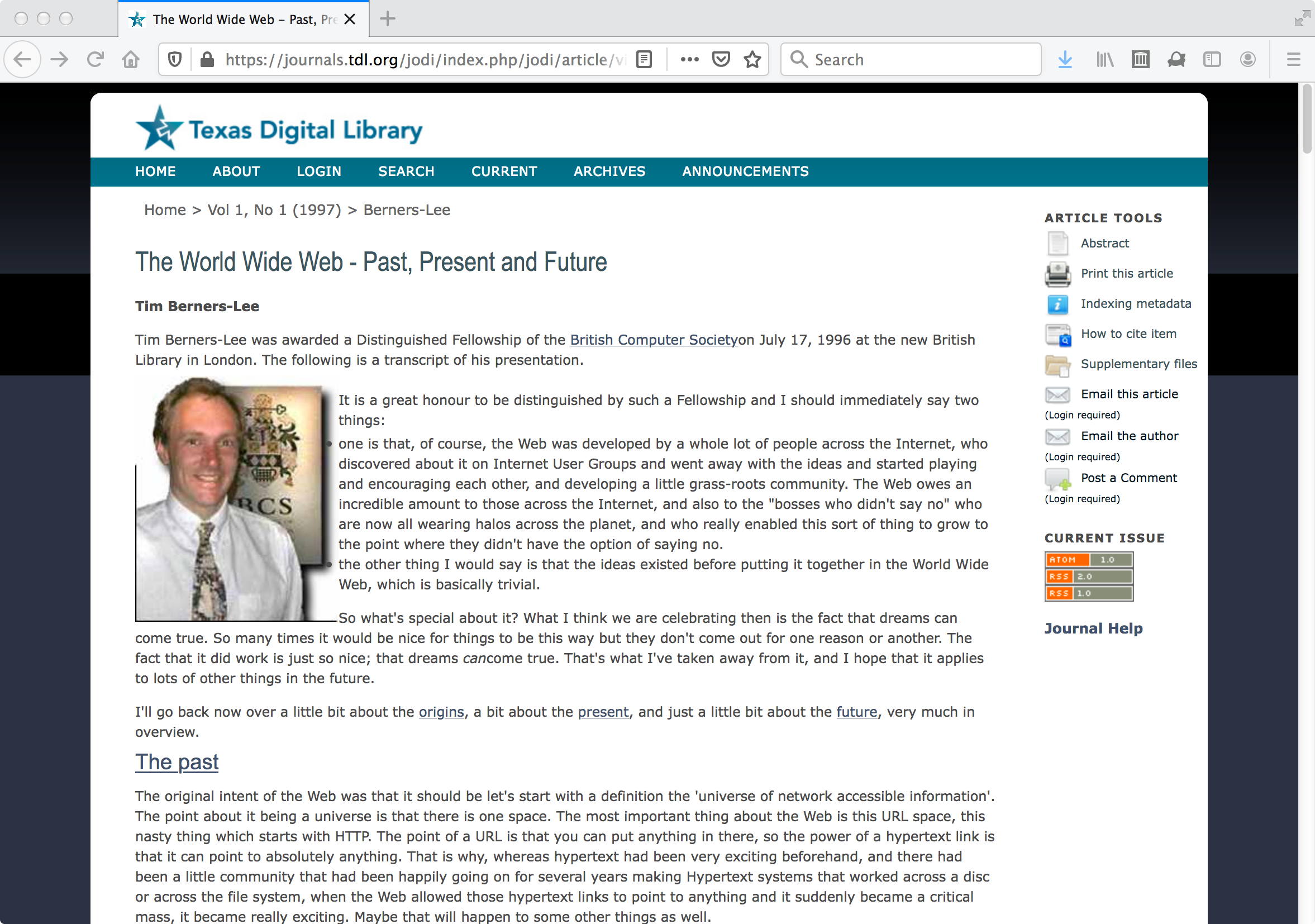}
\caption{The same article on the live web in 2020. \url{https://journals.tdl.org/jodi/index.php/jodi/article/view/3/3}.}
\label{fig:jodi-live}
\end{center}
\end{figure}

The web archives have archived registration walls, since JoDI originally
required a (free) account and login to browse.  Since the Internet
Archive crawls only the surface web (i.e., no login credentials), the
end result is the earliest versions of JoDI were not web archived around
the time they were published.  Eventually the requirement for logins
ceased, and the earliest web archived pages without a registration wall
are from 2000, including snapshots of the earliest articles created two
years after their original publication.  In Figure \ref{fig:jodi-landing-archived}, we see an archived
landing page for an article from the first issue of JoDI (1997). Clicking
through (Figure \ref{fig:jodi-not-archived}) returns a 404 page from the Internet Archive since
that article itself was not archived at that location because of login
restrictions.  Figure \ref{fig:jodi-issue-archived} shows an archived copy of that same article,
meanwhile available at a different URI, created in 2000 when JoDI had
removed the registration wall, and Figure \ref{fig:jodi-live} shows the same article now.  \\

\section{Reflecting on the first issue's articles}
\label{sec:reflecting}

With the vantage point of 25 years, we can properly assess the
significance of the first issue of D-Lib Magazine, especially the
first three articles they published.  Two of the articles introduced
technologies that continue to shape the digital library community
(Dublin Core and DOIs), and the other article is a testament to the
significant funding that the NSF, DARPA, and NASA put into research in
digital libraries, with one of the most prominent outcomes being Google
\cite{nsf:google:dli}.  \\

\subsection{Dublin Core}
\label{sec:dc}

\begin{figure}
\begin{center}
\includegraphics[scale=0.25]{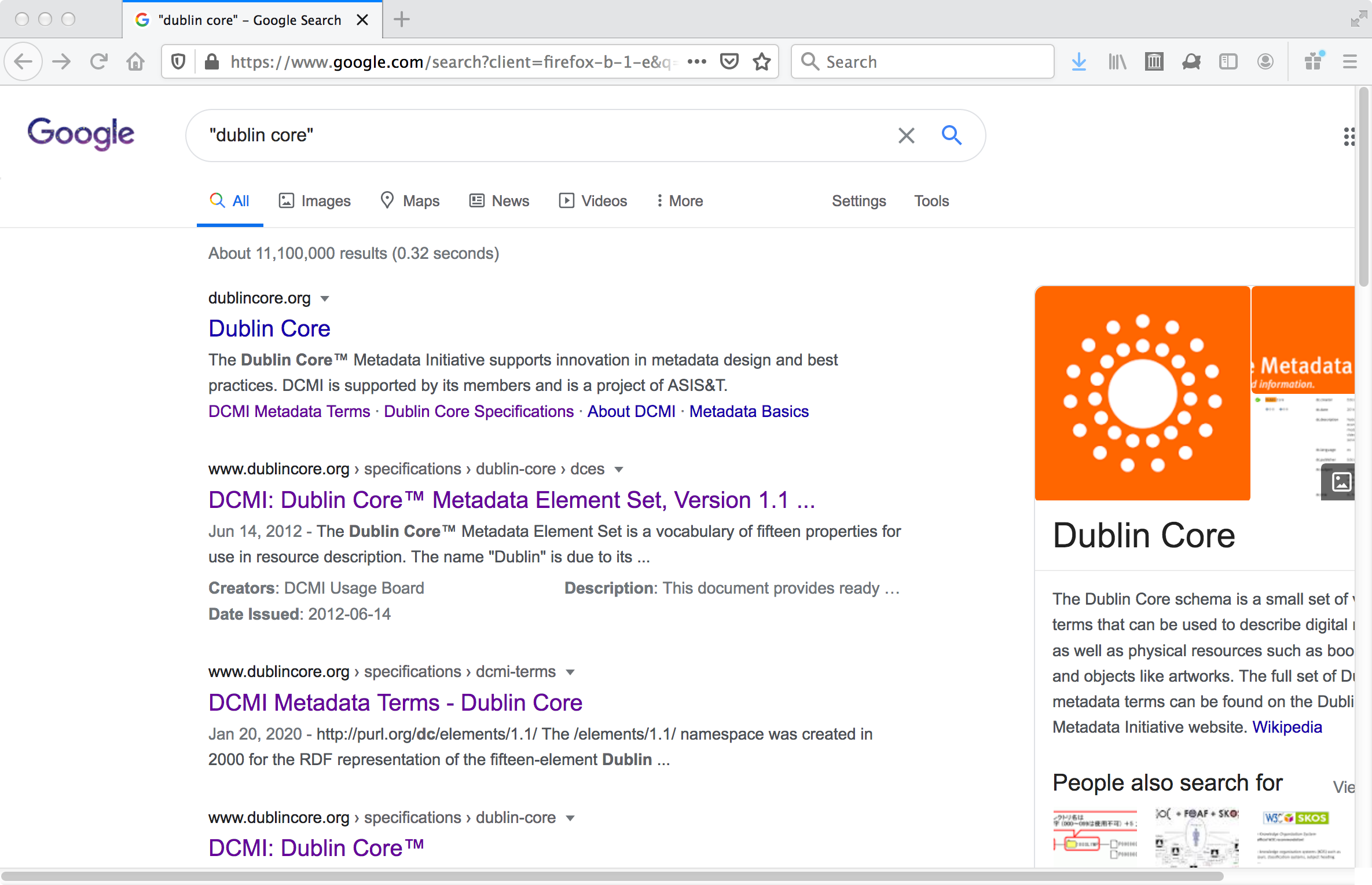}
\caption{11M+ hits for a Google search for ``dublin core''.}
\label{fig:dc-google}
\end{center}
\end{figure}

\begin{figure}
\begin{center}
\includegraphics[scale=0.25]{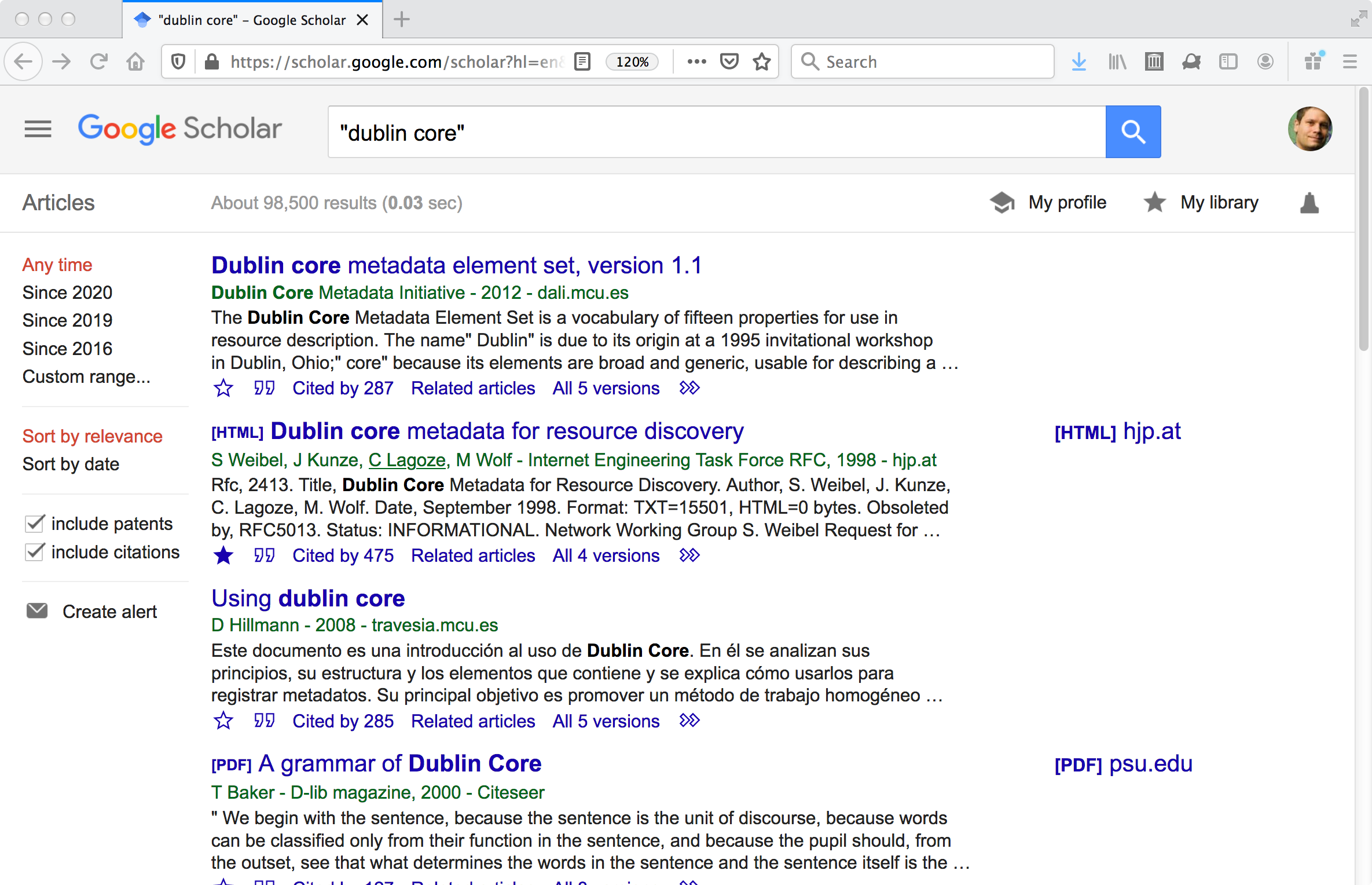}
\caption{98K+ hits for a Google Scholar search for ``dublin core''.}
\label{fig:dc-gs}
\end{center}
\end{figure}

The first article, ``Metadata: The Foundations of Resource Description''
\cite{weibel:dublincore}, is a summary of the OCLC/NCSA Metadata
Workshop Report, which resulted from the workshop in Dublin, Ohio,
only four months prior (March, 1995) \cite{oclc-metdataworkshop}.
The Dublin Core Metadata Element Set (DCMES, or ``Dublin Core'') was
still forming at this point, with only 13 metadata elements, not the
final 15, defined, and ``DCMES'' becoming the Dublin Core Metadata
Initiative (DCMI) Terms.  While the DCMI has gone on to issue over 70
specifications\footnote{\url{https://www.dublincore.org/specifications/dublin-core/}},
today's DCMI Terms can trace their origin to the 1995 Metadata Workshop
and the original DCMES (Table \ref{tab:dc}).  The impact of Dublin Core
is far beyond what we can cover here, but Figure \ref{fig:dc-google}
shows a search for ``dublin core'' in Google yields over 11M hits, and
Figure \ref{fig:dc-gs} shows a similar search in Google Scholar yields
over 98K hits.  \\

\begin{table}[ht]
\caption{Original 1995 DC elements and the current terms.}
\centering
\begin{tabular}{l | l}
\hline\hline
1995 DCMES & Current DCMI Terms \\
\hline
Subject & Subject \\
Title & Title \\
Author & Creator \\
Publisher & Publisher \\
OtherAgent & Contributor \\
Date & Date \\
ObjectType & Type \\
Form & Format \\
Identifier & Identifier \\
Relation & Relation \\
Source & Source \\
Language & Language \\
Coverage & Coverage \\
 & Description \\
 & Rights \\
\end{tabular}
\label{tab:dc}
\end{table}

Dublin Core would form its own community, complete with its
own governance and document series.  But D-Lib Magazine would
continue to be a venue for conveying the status of Dublin
Core \cite{weibel:warwick, lagoze:warwick, weibel:metadataworkshop:1997, thiele:dublincore},
and other related Web metadata efforts, such as PICS
\cite{miller:w3clibraries} and its progeny,
RDF \cite{miller:rdf}, and IEEE
LOM \cite{weibel:lom}.  \\

While Dublin Core is abundantly used for the
description of assets in a variety of content management
systems\footnote{\url{https://lov.linkeddata.es/dataset/lov/}}, continues
to this day to play a role in web-based discovery, co-existing with
similar formats such as the Open Graph Protocol \cite{haugen2010open}
(Figure \ref{fig:lc-curl}) and Schema.org \cite{guha2016schema}
(Figure \ref{fig:datacite-curl}), yet facing some significant
competition from the latter when it comes to Search Engine Optimization
\cite{seopressor:dcschemaorg}.    \\

\begin{figure}
\begin{scriptsize}
\begin{verbatim}
% curl -s https://www.loc.gov/ | grep 'name=\"dc\.\|property=\"og:'
<meta name="dc.identifier"
<meta name="dc.language" content="eng" />
<meta name="dc.source" content="Library of Congress, Washington, D.C. 20540 USA" />
<meta property="og:site_name" content="The Library of Congress"/>
<meta property="og:type" content="article" />
    <meta name="dc.title"
    <meta property="og:title"
        <meta property="og:description" content="The world's largest library. View historic 
        photos, maps, books and more. Contact experts for help with research. Plan a visit. 
        Home of U.S. Copyright Office." />
    <meta name="dc.rights" content="Text is U.S. Government Work" />
<meta property="og:image" content='http://www.loc.gov/static/images/favicons/open-graph-logo.png' />
<meta property="og:image:secure_url" 
  content='https://www.loc.gov/static/images/favicons/open-graph-logo.png' />
<meta property="og:image:width" content="1200"/>
<meta property="og:image:height" content="630"/>
\end{verbatim}
\end{scriptsize}
\caption{The Library of Congress home page with both Dublin Core (\texttt{dc.}) and Open Graph (\texttt{og:}) support.}
\label{fig:lc-curl}
\end{figure}

\begin{figure}
\begin{scriptsize}
\begin{verbatim}
% curl -s https://search.datacite.org/works/10.5281/zenodo.2597274 
<!DOCTYPE html>
<html>
<head>
<title data-conneg='https://api.datacite.org' id='site-title'>
DataCite Search
</title>
<meta content='width=device-width, initial-scale=1.0' name='viewport'>
<link href='//maxcdn.bootstrapcdn.com/font-awesome/4.6.1/css/font-awesome.min.css' 
  rel='stylesheet' type='text/css'>
<link href='//fonts.googleapis.com/css?family=Raleway:400,600,400italic,600italic' 
  rel='stylesheet' type='text/css'>
<link href='//cdnjs.cloudflare.com/ajax/libs/cc-icons/1.2.1/css/cc-icons.min.css' 
  rel='stylesheet' type='text/css'>
<script src='https://unpkg.com/vue/dist/vue.min.js'></script>
<script src='https://unpkg.com/datacite-components/dist/datacite-components.min.js' 
  type='text/javascript'></script>
<link href='https://assets.datacite.org/stylesheets/datacite.css' rel='stylesheet' 
  type='text/css'>
<link href='/stylesheets/usage.css' rel='stylesheet' type='text/css'>
<meta name="DC.identifier" content="10.5281/zenodo.2597274" />
<meta name="DC.type" content="work" />

<meta name="DC.publisher" content="Zenodo" />
<meta name="DC.date" content="2018" />
<script type='application/ld+json'>
{
  "@context": "http://schema.org",
  "@type": "ScholarlyArticle",
  "@id": "https://doi.org/10.5281/zenodo.2597274",
  "identifier": {
    "@type": "PropertyValue",
    "propertyID": "URL",
    "value": "https://zenodo.org/record/2597274"
  },
[deletia]
\end{verbatim}
\end{scriptsize}
\caption{DataCite using both Dublin Core (in the \texttt{meta} elements) and schema.org (in JSON-LD format).}
\label{fig:datacite-curl}
\end{figure}

\subsection{DLI and DLI2}

The second article, ``An Agent-Based Architecture for Digital
Libraries'' \cite{umdl:birmingham}, is a high-level summary of the University
of Michigan Digital Library (UMDL) project, one of the original six
NSF/DARPA/NASA Digital Library Initiative (DLI) projects.  The DLI ran
from 1994--1998, so the 1995 article only summarizes the earliest results. \\

The architectural details of the UMDL are academically interesting, but
the real value in 2020 is reading the article as a time capsule of 1990s
perception of the Web, DLs, and DL architecture.  A quote from near the
beginning of the article describes a scenario that we have since seen
come to pass:

\begin{quote}
The WWW, while it probably contains more information than any single traditional library, is arguably not as useful as a traditional library because it lacks these services (particularly organization and sophisticated search support). No one is dismantling their libraries because of the WWW yet.
\end{quote}

The envisioned architecture focuses heavily on agents, which navigate
a distributed, heterogeneous tapestry of distributed repositories on
behalf of the user.  The model of distributed search was dominant in
early DL architecture thinking, and was reflected in the design of
search protocols like Z39.50 and WAIS, as well as DLs such as WATERS
\cite{maly1995wat}, NCSTRL \cite{DBLP:journals/jasis/DavisL00}, NTRS \cite{ntrs:ir}, and many other examples\footnote{Bill Arms indirectly notes that such architectural decisions trace back to 1991 \cite{arms:2005}, see also Bill Mischo's reflections on federation \cite{mischo:2005}.}.  The DL
commitment to distributed searching on the Web culminated in the STARTS
protocol \cite{gravano1997ssp}, and dissatisfaction with the state of distributed
searching DLs (cf. \cite{powell2000gas, nelson:object} was at the heart
of the Universal Preprint Service prototype that demonstrated metadata
harvesting and centralized searching, a design decision that would
inform OAI-PMH \cite{oaipmh:ups,oaipmh:sfc,oaipmh:protocol} and later DLs based on it (e.g.,
\cite{liu2001aos,544256,996466}).  \\

Typical of the time, the UMDL design is fully committed to distributed
search and crawling, with personalized agents handling the foraging and
negotiation with the various repositories (similar to CNRI's Knowbots
\cite{knowbots}).  After 25 years and with a post-Google perspective, we can
now see that most meta-search / distributed search architectures have
been retired, in part by the hegemony of Google-style crawling and
searching\footnote{Unified Computer Science Technical Report Index
(UCSTRI) \cite{ucstri}, a computer science DL that independently
crawled and indexed anonymous FTP sites is the first known example of the
architecture that successful DLs like CiteSeer \cite{giles1998citeseer}
and Google Scholar \cite{verstak2014shoulders} would employ.}.  In the
end, (logically) centralized architectures won.  HTTP servers were (and
are) broadly distributed, but the complexity of crawling and indexing
turned out to be centralized.  The search engines now dictate to the web
servers how to expose and structure their site, instead of the anticipated
model where sites instructed the best way to access their holdings.   \\

The DLI ran from 1994 through 1998, and its \$30M supported six projects\footnote{\url{https://web.archive.org/web/19981202064413/http://www.cise.nsf.gov/iis/dli\_home.html}},
each of which is summarized in the one year anniversary issue (July/August 1996)
of D-Lib Magazine\footnote{\url{http://hdl.handle.net/cnri.dlib/july96}}:

\begin{itemize}
\item ``The University of Michigan Digital Libraries Research Project'' led by the University of Michigan \cite{dlib:dli:um}

\item ``Building the Interspace: Digital Library Infrastructure for a University Engineering Community'' led by the University of Illinois \cite{dlib:dli:uiuc}

\item ``The Environmental Electronic Library: A Prototype of a Scalable, Intelligent, Distributed Electronic Library'' led by the University of California, Berkeley \cite{dlib:dli:berkeley}

\item ``Informedia: Integrated Speech, Image and Language Understanding for Creation and Exploration of Digital Video Libraries'' led by Carnegie Mellon University \cite{dlib:dli:cmu}

\item ``The Stanford Integrated Digital Library Project'' led by Stanford University \cite{dlib:dli:stanford}

\item ``The Alexandria Project: Towards a Distributed Digital Library with Comprehensive Services for Images and Spatially Referenced Information'' led by the University of California, Santa Barbara \cite{dlib:dli:ucsb}

\end{itemize}

DLI2 ran from 1999--2005, and the \$55M from the NSF, DARPA, National
Library of Medicine, Library of Congress, NASA, and the National
Endowment for the Humanities (with additional participation from the
National Archives and the Smithsonian Institution) supported 36 projects.
Despite the importance of the DLI and DLI2 funding efforts in the early
days of the web, very little about the funding programs remains on the live
web outside of what is hosted at dlib.org.  Sites like \url{www.dli2.nsf.gov}
and \url{www.cise.nsf.gov/iis/dli\_home.html} are no longer on the live web,
with only a single page left at nsf.gov to mark 12 years of research
and \$85M in funding\footnote{\url{https://www.nsf.gov/news/news\_summ.jsp?cntn\_id=103048}}. Although the former pages are accessible in web
archives\footnote{For example, \url{https://web.archive.org/web/*/http://www.cise.nsf.gov/iis/dli\_home.html} and \url{http://web.archive.org/web/*/http://www.dli2.nsf.gov/projects.html}} and the individual, specific technical contributions resulting
from the DLI work are widely described in the broader literature, D-Lib
Magazine was a prime venue for program-level reflection 
\cite{griffin:1998, griffin:1999, lesk:1999, hirtle:1999, arms:2000:editorial, larsen:2005, griffin:2005, paepcke:2005}.\\

\subsection{KWF and DOIs}
\label{sec:kwf}

The final article from the first issue, ``Key Concepts in the Architecture
of the Digital Library'' \cite{kwf:arms},  was a summary by Bill Arms of ``A framework
for distributed digital object services'', which would later be
known as the ``Kahn-Wilensky Framework'' (KWF) \cite{kwf}.
Although it is just an abstract framework and not tied to a specific
implementation, the KWF has had a significant impact on the architectural
design of digital libraries, especially concerning the identification
and structure of `digital objects'' and their relationship with the
repositories in which they reside.  In 2006, we edited a special issue
of the International Journal on Digital Libraries (IJDL) on ``Complex
Digital Objects'' \cite{intro:co:ijdl}, which also featured a reprint
and of the KWF along with commentary from Robert Kahn \cite{kwf:ijdl}.  \\

The KWF provided the architecture for the initial CS-TR project
\cite{cs-tr:kahn}, which in combination with WATERS \cite{maly1995wat} formed the
basis for the Dienst protocol and NCSTRL \cite{DBLP:journals/jasis/DavisL00} as a distributed
digital library for computer science technical reports.  Lessons learned
from Dienst were incorporated in OAI-PMH \cite{oaipmh:ups,oaipmh:sfc,oaipmh:protocol}.
The KWF also had an impact in the Dublin Core community, resulting in
the Warwick Framework \cite{866869}, which was later extended with
``distributed active relationships'' \cite{danieljr1997ewf}, which itself later
evolved into Fedora \cite{696688}.  The management of Fedora and DSpace \cite{827151}
were merged into Duraspace in 2009 \cite{duraspace:2009:tripreport}, and in 2019 LYRASIS absorbed
Duraspace. The separate Fedora and DSpace open source products continue 
to be offered. \\

KWF specified the role of repositories in mediating access
to their digital objects via the Repository Access Protocol
(RAP).  Over the years, numerous papers have been published,
many of them in D-Lib Magazine, that pertain to RAP, including
\cite{arms1997aid, cornell:cnri, reilly:2010}.
The design
of RAP was repository-centric \cite{dlib:2015:reminiscing} and explicitly
decoupled the protocol for expressing interactions with digital
objects from the transport protocol used to transfer interaction
requests between client and server. Such a choice was not uncommon in
the days preceding the dominance of the web and its now omnipresent
HTTP protocol and can, for example, also be observed in the design
of OAI-PMH \cite{dlib:2015:reminiscing}.  Many transport protocols (TCP,
SMTP, FTP, Gopher, HTTP, IIOP, etc.) overlapped in time and there was
a predisposition to viewing them as impermanent, interchangeable;
something on which one built richer, domain-specific protocols. Also,
Roy Fielding did not publish his dissertation about Representational
State Transfer (REST), which made the resource-centric \cite{dlib:2015:reminiscing}
semantics and potential of HTTP explicit, until 2000 \cite{fielding,
fielding2002principled}. By that time, the state of thinking and practice
in digital libraries had already diverged from that of the web. In
this way, RAP was the initial manifestation of an architectural
fault line between digital libraries and the web that continues
to this day\footnote{An excellent review of the complex relationship between DLs and the Web is Carl Lagoze's 2010 dissertation, ``Lost Identity: The Assimilation Of Digital Libraries Into The Web'' \cite{lagoze}.}. As a prominent example, the FAIR Digital Object effort
\cite{de2020fair} that fits under the broad
umbrella of the European Open Science Cloud\footnote{\url{https://www.eosc-portal.eu/}} program and is supported
by activities of the Research Data Alliance\footnote{\url{https://www.rd-alliance.org/}}, considers two approaches\footnote{\url{https://github.com/GEDE-RDA-Europe/GEDE/tree/master/FAIR Digital Objects/FDOF}}
to devise rich interactions with digital objects that are stored
in cooperating repositories. One aligns with the RAP line of thought
\cite{dona:doip}.
The other embraces a webby approach and advocates leveraging a range
of HTTP-based standards that have become available over the years,
including the Open Archives Initiative Object Reuse and Exchange
(OAI-ORE) \cite{Lagoze:2008th} for the representation of digital objects as
aggregations of web resources,  the Memento Framework \cite{nelson:memento:tr,
memento:rfc} for temporal versioning of web resources,  Linked Data
Platform \cite{ldp:w3c} and the Fedora API
\cite{fedora:api} for CRUD
operations on digital objects and their constituent resources,
Web Annotation \cite{w3c:annotation:model, w3c:annotation:protocol}, RO-Crate for packaging
digital objects \cite{rocrate}. It is
interesting to note that several of these specifications were co-authored
by people with roots in the Digital Library community.   \\

The most visible contribution from the KWF is Digital Object
Identifiers (DOIs).  Handles \cite{lannom:handles}, of which DOIs are
a subset, were part of the technical infrastructure for digital
libraries built by CNRI. Although frequently considered a URN
implementation \cite{rfc:8141}, they are not registered as URN namespaces\footnote{\url{https://www.iana.org/assignments/urn-namespaces/urn-namespaces.xhtml}}
and their status as URIs remains unresolved \cite{rfc:3652}.   D-Lib
Magazine used handles beginning with the first issue, and, as the DOI
effort matured, D-Lib Magazine was naturally an early adopter, starting
in January, 1999 \cite{wilson:1999:editorial}.  Figure \ref{fig:curl-handle}
shows a current resolution of a handle from Arms's description of KWF
(\texttt{hdl:cnri.dlib/july95-arms}), and Figure \ref{fig:curl-doi}
shows the resolution of that handle converted to DOI format
(\texttt{doi:10.1045/july95-arms}, with: \texttt{hdl} $\rightarrow$ 
\texttt{doi} and \texttt{cnri.dlib} $\rightarrow$ \texttt{10.1045}). 
In fact, all DOIs are also resolvable as handles, as Figure \ref{fig:dois-handles} shows.  
But since DOIs are a proper subset of handles, not all handles are resolvable as DOIs (Figure \ref{fig:handles-dois}).  \\

\begin{figure}
\begin{scriptsize}
\begin{verbatim}
$ curl -I http://hdl.handle.net/cnri.dlib/july95-arms
HTTP/1.1 302
Location: http://www.dlib.org/dlib/July95/07arms.html
Expires: Tue, 21 Jul 2020 15:30:58 GMT
Content-Type: text/html;charset=utf-8
Content-Length: 171
Date: Mon, 20 Jul 2020 15:30:58 GMT
\end{verbatim}
\end{scriptsize}
\caption{Resolution of a handle from the first issue of D-Lib Magazine.}
\label{fig:curl-handle}
\end{figure}

\begin{figure}
\begin{scriptsize}
\begin{verbatim}
% curl -IL https://doi.org/10.1045/july95-arms
HTTP/2 302
date: Mon, 20 Jul 2020 15:55:01 GMT
content-type: text/html;charset=utf-8
content-length: 171
set-cookie: __cfduid=da38cfa9ab1b68408e00ad2c5d8c678541595260501; 
  expires=Wed, 19-Aug-20 15:55:01 GMT; path=/; domain=.doi.org; HttpOnly; 
  SameSite=Lax; Secure
vary: Accept
location: http://www.dlib.org/dlib/July95/07arms.html
expires: Mon, 20 Jul 2020 16:25:51 GMT
cf-cache-status: DYNAMIC
cf-request-id: 040e8894eb0000f11e27b76200000001
expect-ct: max-age=604800, 
  report-uri="https://report-uri.cloudflare.com/cdn-cgi/beacon/expect-ct"
strict-transport-security: max-age=31536000; includeSubDomains; preload
server: cloudflare
cf-ray: 5b5ddd34add1f11e-IAD
\end{verbatim}
\end{scriptsize}
\caption{Resolution of a DOI formed from the handle shown in Figure \ref{fig:curl-handle}.} 
\label{fig:curl-doi}
\end{figure}

\begin{figure}
\begin{scriptsize}
\begin{verbatim}
% curl -I https://doi.org/10.1002/cpe.1594
HTTP/2 302
date: Fri, 24 Jul 2020 17:14:50 GMT
content-type: text/html;charset=utf-8
content-length: 159
set-cookie: __cfduid=d3e73c3df617628234c082278fdcc7f1a1595610890; 
  expires=Sun, 23-Aug-20 17:14:50 GMT; path=/; domain=.doi.org; HttpOnly; 
  SameSite=Lax; Secure
vary: Accept
location: http://doi.wiley.com/10.1002/cpe.1594
expires: Fri, 24 Jul 2020 17:20:12 GMT
cf-cache-status: DYNAMIC
cf-request-id: 04236b1a080000031614326200000001
expect-ct: max-age=604800, 
  report-uri="https://report-uri.cloudflare.com/cdn-cgi/beacon/expect-ct"
strict-transport-security: max-age=31536000; includeSubDomains; preload
server: cloudflare
cf-ray: 5b7f47a34c260316-IAD

% curl -I http://hdl.handle.net/10.1002/cpe.1594
HTTP/1.1 302
Vary: Accept
Location: http://doi.wiley.com/10.1002/cpe.1594
Expires: Fri, 24 Jul 2020 18:03:23 GMT
Content-Type: text/html;charset=utf-8
Content-Length: 159
Date: Fri, 24 Jul 2020 17:15:08 GMT
\end{verbatim}
\end{scriptsize}
\caption{All DOIs are also handles.}
\label{fig:dois-handles}
\end{figure}

\begin{figure}
\begin{scriptsize}
\begin{verbatim}
% curl -I https://www.doi.org/cnri.dlib/july95-arms
HTTP/1.1 404 Not Found
Content-Type: text/html
Content-Length: 3065
Connection: keep-alive
Last-Modified: Wed, 01 Mar 2017 01:16:05 GMT
Server: AmazonS3
Date: Thu, 06 Aug 2020 13:28:05 GMT
ETag: "39bf1abd89479be3047e0cc48f631b42"
Vary: Accept-Encoding
X-Cache: Error from cloudfront
Via: 1.1 6784ac36b8d920a78daf15294a50025f.cloudfront.net (CloudFront)
X-Amz-Cf-Pop: IAD79-C3
X-Amz-Cf-Id: oPtjDhhUnKYjw4pCALyY5ECCPcQeuRd5lyl40T6mUTzd5HefGSnIWQ==
Age: 1988
\end{verbatim}
\end{scriptsize}
\caption{Not all handles are DOIs.}
\label{fig:handles-dois}
\end{figure}

\FloatBarrier

\section{Progress over 25 years}
\label{sec:progress}

The last part of the first issue we would like to review is a page
entitled ``To the editor: What's needed in future research?''\footnote{\url{http://www.dlib.org/dlib/July95/07messages.html}}, in
which the D-Lib editors polled five prominent digital library researchers
and administrators and asked them to briefly identify and discuss areas
that warranted further research.  Although the purpose was to generate
discussion about a near-term research agenda (and perhaps establish D-Lib
Magazine's credentials through adding additional voices to the first
issue), this page now serves as a time capsule and allows us to reassess
progress in the field since 1995.  In the page the editors stated ``Now,
we would like to know what you think; send [us] your thoughts, reactions,
and comments'', which we now do 25 years later. 

\subsubsection*{``An interoperable national and global information web'' --  Barry M. Leiner, Deputy Director, ARPA/CSTO}

The world-wide web technology, infrastructure, and protocols that Barry
Leiner credited for an explosion in the availability of accessible and
viewable information have persisted and have given rise to a global
networked environment that we can hardly imagine living without
anymore. Over time, numerous open standards have been specified to
support interoperability beyond the basic level provided by the core
ingredients of the web, HTTP and HTML. In the Web 2.0 era, some of
these acted as catalysts for the frictionless creation of value-added
services across web platforms, both for and not for profit. But since
interoperability is not a significant concern for companies that want
to protect their turf or establish monopolies, a trend has emerged to
support rich access by means of bespoke APIs rather than open protocols,
significantly increasing the investment for the development and management
of services that require cross-platform interactions. This trend has
become so prominent that platforms routinely claim to be interoperable
because they expose a self-defined API, and, to add insult to injury,
touting its RESTful-ness while many times it is not \cite{fielding:apis}. \\

In the digital library community, the dream to achieve interoperability
based on open standards remains alive and actively pursued.  Despite
the aforementioned ongoing debate regarding which path to take --
repository-centric or resource-centric -- most community driven
specification efforts of the past decade have chosen to embrace
the ways of the web, many times aiming for approaches that have
applicability beyond the digital library community. In the realm of
technologies to support digital libraries of multimedia information on
which Barry Leiner zoomed in, prominent examples include the Fedora
API \cite{fedora:api} that leverages the W3C Linked Data Platform
recommendation \cite{ldp:w3c}, the W3C Web Annotation recommendations
\cite{w3c:annotation:model, w3c:annotation:protocol}, and the
specifications that resort under the International Image Interoperability
Framework\footnote{\url{https://iiif.io/technical-details/}}. Because of
its growing global adoption by GLAM institutions, especially the latter
stands as a testimony that rich interoperability for distributed resource
collections is effectively achievable. But other promising specifications
that aim for the same holy grail are struggling for adoption, and,
many times, lack of resources is mentioned as a reason. While that
undoubtedly plays a role, it did not stand in the way of rapid
adoption of protocols that have emerged from large corporations,
such as the Google-dominated \cite{google:dominate:schemaorg}
schema.org\footnote{\url{https://schema.org}}. This consideration
re-emphasizes that a core ingredient of a successful interoperability
specification, and hence of achieving an interoperable global information
web, is a large megaphone, either in the guise of commercial power or
active community engagement \cite{8791114}.  

\subsubsection*{``Integration between electronic and non-electronic forms of communications and publications'' -- Ann L. Okerson, Director, Office of Scientific \& Academic Publishing, Association of Research Libraries (ARL)}  

The dichotomy between print and digital resources that Ann Okerson
describes was a major concern in the mid nineties. Despite large-scale
digitization efforts (see below) and the exponential growth of
born-digital materials, analog collections will remain. But these
worlds are no longer perceived as being radically distinct because
the analog world has largely been absorbed by the digital one. This
did not necessarily happen by making both discoverable through library
OPACs. Rather, it has become commonplace to cater to the crawl-driven
discovery paradigm of major search engines by exposing resource
descriptions for materials of both types of collections to the web using
Search Engine Optimization techniques such as the Sitemap protocol and,
more recently, schema.org. For many analog GLAM collections doing so
requires making traditional catalog systems web savvy, a task that
is still ongoing.  So while ``electronic and non-electronic forms of
communications and publications'' will remain parallel worlds for
the foreseeable future, the percentages have shifted, as Lannom's
note from the final D-Lib editorial makes clear: ``the very phrase
`Digital Libraries' has gone from sounding innovative to sounding a bit
redundant.'' \cite{lannom:2017:editorial}  

\subsubsection*{```Foreground' information stores, or personal digital libraries'' -- William L. Scherlis, Senior Research Computer Scientist, Department of Computer Science, Carnegie Mellon University}

The personal digital libraries envisioned in this piece have not
become mainstream. Instead creators have embraced a myriad of web
productivity portals to share both their intellectual artifacts and
daily activities. As a result, assets created by individuals and
information pertaining to their comings and goings are distributed
across the web, to such an extent that both research (e.g.,
\cite{guy}) and development\footnote{for example: \url{https://github.com/LockerProject/Locker}} efforts have considered
approaches to aggregate it into a personal environment that provides a
concise representation of the self on the distributed web. In the realm
of scholarly communication, the experimental myresearch.institute\footnote{\url{https://myresearch.institute}} effort
tracks, collects, and archives assets created by researchers in a variety
of web portals, including GitHub, Slideshare, Wikipedia. The plethora of
APIs used by these portals and the lack of support for protocols such as
W3C ActivityPub\cite{w3c:activitypub} and W3C ActivityStreams2 \cite{w3c:activitystreams2} make such an aggregation task
far from trivial. The result of gathering the distributed information
could be considered a proxy personal digital library. But maybe the days
of the actual personal digital library are still to come. Motivated and
frustrated by the monopolies certain web portals have established over
the years, and the concerns regarding privacy and data abuse that result,
the Decentralized Web movement is aiming for alternatives, with a focus
on giving individuals back control over their personal assets. As part
of this movement, the Solid\footnote{\url{https://inrupt.com/solid}} effort led by Tim Berners-Lee introduces the
notion of a pod \cite{tbl:cloud},
a personal storage space that complies with a stack of open standards and
allows its owner to grant or deny applications and users access rights
to stored resources. Clearly, these pods are conceived as a technology
that can help information producers manage foreground information,
to put it in the words of William L. Scherlis.  

\subsubsection*{``Diversifying and access'' and ``the distribution chain'' -- Paul Evan Peters, Executive Director, Coalition for Networked Information} 

As far as ``diversifying and access'', we believe Peters is
criticizing researchers' emphasis on the ``attributes of the
resources and services'', as manifested in Leiner's assessment, as
well as the prevalent distributed searching paradigm described in section
\ref{sec:kwf}, and instead we should be focusing on the ``attributes
of the users and uses'', which we interpret to be in harmony with
Scherlis's vision of personal DLs.  Most siloed repositories have
been flattened and exposed for crawling by search engines.  Google, for
example, showed little appetite for even simple protocols such as OAI-PMH
\cite{mccown:search, hagedorn:2008},
and officially retired support for it in 2008
\cite{google:pmh}. \\

Initially we struggled with understanding Peters's description
of ``the distribution chain.'' Eventually we decided that part of
the ambiguity is that he is describing something that is so common
now but for which the language did not exist in 1995, resulting in
a terminology gap that we had to bridge to understand what he meant.
When he says we need to focus on ``closing the gap between creators
and users of resources and services'', we understand that to be
an admonishment that the point of DLs should not be merely the
automation of the existing publication process. Unfortunately, 2020
still resembles 1995, with the distribution chain largely paralleling
the value chain, just now with PDFs instead of paper.  Instead of
reenvisioning / reengineering the scholarly communication process (e.g.,
\cite{Van-de-Sompel:2004zr}),
we have a confusing array of open access options (``gold'',
``green'', ``hybrid'', etc.\footnote{\url{https://en.wikipedia.org/wiki/Open\_access}}), and by retaining the
publisher at the center of distribution, they still fail to
address the broader needs of scholarly communication (e.g.,
\cite{brembs:scholarship}).\\

We have now long had the ability to ``link creators and users'' as
Peters's calls for, but have lacked the collective will to make the
transition \cite{capadisli, vandesompel:pods:2017}.  Web 2.0, blogs,
and social media provided some hope initially, but as noted in section
\ref{sec:kwf}, platforms have since moved away from Atom and RSS in
favor of bespoke APIs that provide more functionality at the expense
of interoperability.  Add to this Elsevier buying various platforms
that ``link creators and users'' (e.g., SSRN \cite{pike2016elsevier},
Mendeley \cite{elsevier:mendely}, bepress \cite{elsevier:bepress},
and Peters's vision, while technically feasible, seems no closer than
it did in 1995. 

\subsubsection*{``Retrospective capture of content'' -- James Michalko, President, The Research Libraries Group, Inc.} 

Michalko, in a statement that aligns with Okerson's, observes
``[t]here's a major opportunity and demand for the retrospective
capture of content'', but in 1995 ``[t]here are few service bureaus
that can do the scanning and capture of maps, manuscripts, and other
primary research materials.''  However, Michalko was writing at a time
when these projects either did not exist or were just beginning: JSTOR
\cite{schonfeld2012jstor}, Google Books \cite{googlebooks:blog}, Open Content Alliance \cite{opencontentalliance:kahle}, Internet
Archive \cite{ia:openlibrary:2010}, HathiTrust \cite{york2009library}, National Digital Newspaper Program
\cite{aguera2006us}, museum mass digitization projects \cite{primary2014survey}, etc.
Mass digitization of primary research materials remains incomplete,
and it is not always clear how the digitized models will fit within a
search engine-centric model of crawling and searching.  But the momentum
is there, and what is digitized exists at a scale that we could only
dream of in 1995. \\ 

\section{Conclusions}
\label{sec:conclusions}

D-Lib Magazine was published from 1995 through 2017.  During this time,
it helped shape the digital library community, via the information
published in articles, the ancillary awareness and informational updates
now largely provided by social media, and as an ongoing experiment in
web(-only) publishing.  Although it ceased publication three years ago,
the entire site is still on the live web as an unchanged time capsule,
and as a serial it still accrues many citations. The 1,062 articles and
5,000+ web pages available at www.dlib.org offer many opportunities for
reflection about the DL community, but we took the first issue as our
point of reference to review the last 25 years. \\

Of the three articles in the first issue, all were summaries of work
described elsewhere. However, only the article about the DLI-funded
UMDL project summarized information in conventional, peer-reviewed
publications.  The other two articles, about Dublin Core and the
Kahn-Wilensky Framework, summarized ``unpublished'' (i.e., grey
literature) reports, providing a more formal and citable surrogate for
standards, practitioner, engineering work that was crucial in the early
days of DLs, which the conventional, peer-reviewed publication venues
would largely ignore. \\

D-Lib Magazine would cease publishing due in part to an unsustainable
funding model, the maturity of the field, and the rise of blogs and
social media. However, its role in shaping the then emerging DL community
is hard to overstate.  Given the perspective of 25 years, one would be
hard pressed to retroactively construct a more prescient first issue:
Dublin Core is an ongoing initiative and suite of standards that DLs and
the general web still employ today, the \$30M from the DLI bootstrapped
the DL community and eventually gave us Google, and the Kahn-Wilensky
Framework have influenced the design of repositories (e.g., Fedora),
interoperability (e.g., OAI-PMH), and provided the proof-of-concept
to help launch the DOI ecosystem that provides a fundamental level of
interoperability across scholarly publishing.  \\

\section*{Acknowledgments}
We would like to thank everyone who contributed to D-Lib Magazine's
success, first of which is Robert Kahn, who applied the resources of
CNRI (and DARPA) to make D-Lib Magazine possible.  We appreciate the
work of all the editors through the years: Bill Arms, Amy Friedlander,
Barry Leiner, Bonnie Wilson, Peter Hirtle, Allison Powell, Larry Lannom,
and Cathy Rey.  \\

We are grateful that D-Lib Magazine existed during the maturation of the
field of digital libraries as well as during the pivotal points in our
respective careers.  Although we are happy to publish this paper as both
an arXiv eprint and a blog post, we note with a tinge of sadness that
this paper would have been best suited for publication in D-Lib Magazine.

%
\bibliographystyle{abbrvurl}
\bibliography{mln}  
%

\end{document}